\documentclass[fleqn,usenatbib]{mnras}

\usepackage{newtxtext,newtxmath}

\usepackage[T1]{fontenc}

\DeclareRobustCommand{\VAN}[3]{#2}
\let\VANthebibliography\thebibliography
\def\thebibliography{\DeclareRobustCommand{\VAN}[3]{##3}\VANthebibliography}

\usepackage{graphicx}	
\usepackage{amsmath}	
\usepackage{amssymb}	
\usepackage{siunitx}
\usepackage{comment}
\usepackage{multirow}
\usepackage{physics}
\usepackage{threeparttable}

\title[MAXI~J1820+070 spectral-timing]
{MAXI~J1820+070 X-ray spectral-timing reveals the nature of the accretion flow in black hole binaries}

\author[T. Kawamura et al.]{
Tenyo Kawamura,$^{1,2}$\thanks{E-mail: tenyo.kawamura@ipmu.jp}
Chris Done,$^{3,2}$
Magnus Axelsson$^{4,5}$\thanks{Deceased}
and Tadayuki Takahashi$^{2,1}$
\\
$^{1}$Department of Physics, University of Tokyo, Bunkyo, Tokyo 113-0033, Japan\\
$^{2}$Kavli Institute for the Physics and Mathematics of the Universe (WPI), University of Tokyo, Kashiwa, Chiba 277-8583, Japan\\
$^{3}$Department of Physics, University of Durham, South Road, Durham DH1 3LE, UK\\
$^{4}$Oskar Klein Center for CosmoParticle Physics, Department of Physics, Stockholm University, SE-10691 Stockholm, Sweden\\
$^{5}$Department of Astronomy, Stockholm University, SE-10691 Stockholm, Sweden
}

\date{Accepted XXX. Received YYY; in original form ZZZ}

\pubyear{2022}

\begin{document}
\label{firstpage}
\pagerange{\pageref{firstpage}--\pageref{lastpage}}
\maketitle

\begin{abstract}
Black hole X-ray binaries display significant stochastic variability on short time-scales 
(0.01--100 seconds), with a complex pattern of lags in correlated variability seen in different energy bands. 
This behaviour is generally interpreted in a model where slow fluctuations stirred up at large radii propagate down through the accretion flow, modulating faster fluctuations generated at smaller radii. Coupling this scenario with radially-stratified emission opens the way to measure the propagation time-scale from data, allowing direct tests of the accretion flow structure. We previously developed a model based on this picture and showed that it could fit {\it NICER} (0.5--10~keV) data from the brightest recent black hole transient, MAXI~J1820+070. However, here we show it fails when extrapolated to higher energy variability data from {\it Insight-HXMT}. We extend our model so that the spectrum emitted at each radius changes shape in response to fluctuations (pivoting) rather than just changing normalisation. This gives the strong suppression of fractional variability as a function of energy seen in the data. The derived propagation time-scale is slower than predicted by a magnetically arrested disc (MAD), despite this system showing a strong jet. Our new model jointly fits the spectrum and variability up to 50~keV, though still cannot match all the data above this. Nonetheless, the good fit from 3--40~keV means the QPO can most easily be explained as an extrinsic modulation of the flow, such as produced in Lense-Thirring precession, rather than arising in an additional spectral-timing component such as the jet. 
\end{abstract}




\section{Introduction}
\label{sec:intro}

The nature and geometry of the X-ray emission region in black hole binaries are still controversial, especially in the low/hard state, where most 
of the power is emitted in a spectrum quite unlike a standard disc (\citealt{Shakura_1973}). 
Spectral fitting alone is degenerate, with proposed geometries being a compact source on the spin axis (lamppost), extended emission along the jet direction (jet corona), extended coronal emission on top of an underlying accretion disc (sandwich), and extended coronal emission which replaces the accretion disc (truncated disc/hot inner flow) see e.g.  \cite{Poutanen_2018}.
The truncated disc/hot inner flow model has the advantage that it gives a framework to explain the evolution of the spectrum and its fast variability properties together (\citealt{Done_2007}), although there are persistent questions over the extent of disc truncation from modelling the reflected emission and its associated iron line (e.g. \citealt{Buisson_2019}; but see \citealt{Zdziarski_2021}). 
Another way to track the extent of the disc is the quasi-thermal emission arising from the same X-ray irradiation of the disc, which gives rise to the iron line and reflected emission (\citealt{DeMarco_2015, Wang_2022}). 
Photons which are not reflected are reprocessed in the disc, producing a thermal reverberation signal. This gives a soft lag, where variations of soft photons follow those of hard photons with a light travel time delay.
Reverberation size-scales do indeed point to a truncated disc, with a truncation radius which decreases as the source spectrum softens (\citealt{DeMarco_2021}). 
Perhaps the most compelling evidence for a truncated disc is the new polarisation results for the low/hard state of Cyg~X-1. These rule out the X-ray emission region being aligned with the jet and, instead, require it to be aligned with the accretion flow (\citealt{Krawczynski_2022}). 
Truncated disc/hot inner flow models are thus strongly favoured, motivating our work in exploring how we can derive the physical properties of the hot flow.

The fast variability (0.01--100 seconds) gives independent constraints on the accretion flow. It shows many complex properties which change as a function of energy and variability time-scale (see e.g. the review by  \citealt{Uttley_2014}). The
most promising framework in which to explain these is with propagating fluctuations (\citealt{Lyubarskii_1997, Kotov_2001}). The idea is that variability is generated in the accretion flow (e.g. by the turbulent dynamo magnetorotational instability \citealt{Balbus_1991}), with a characteristic time-scale which is shorter at smaller radii. 
Fluctuations generated at any radius in the hot flow propagate down so that slower fluctuations stirred up at larger radii propagate down to modulate the faster fluctuations produced at smaller radii. This produces correlated but lagged multi-time-scale variability in the entire hot flow (\citealt{Lyubarskii_1997, Kotov_2001}). 
These lags can be seen directly in the data if the flow produces different spectra at different radii, with the observed `hard lags' (fluctuations at 10--20~keV lagging behind the same fluctuation at 2--3~keV; \citealt{Miyamoto_1988,Nowak_1999}) requiring that smaller radii have harder spectra, which seems physically intuitive. Numerical models combining the propagating fluctuations process with a spectrally-inhomogeneous hot flow have shown general agreement with the variability properties observed by {\it RXTE} in the 3-30~keV bandpass (\citealt{Arevalo_2006,Ingram_2011,Ingram_2012}). 

However, accurately reproducing all the observed timing properties with the propagating fluctuations model turned out to be more 
difficult than expected in these first quantitative models, which considered only the variability in the hot flow. 
One approach is simply to bypass this complexity and use phenomenological models of the intrinsic variability and its lags. This is especially useful if the 
goal is simply to measure reverberation lags as in the {\tt RELTRANS} code (\citealt{Ingram_2019b}). However, our goal is instead to make a physical model of the hot flow, self-consistently producing its spectrum and variability. Such models can then be used to measure physical properties e.g. the propagation speed, which constrains the nature and geometry of the hot flow. Hence we extend the hot flow propagating fluctuation models to get a better match to the data. There are many potential ways to do this, but the goal is to identify the physical processes which make the most impact on the observed data.

One difficulty with models of propagating fluctuations through the hot flow is that this typically gives rise to a single-peaked power spectrum (\citealt{Arevalo_2006,Ingram_2011,Ingram_2012}), whereas the observed
power spectra are often double-peaked (\citealt{Belloni_2002, Pottschmidt_2003, Axelsson_2005, Grinberg_2014}).
It is possible to change the time-scales and amplitude of variability with radius in the hot flow to match the data, but it seems fine-tuned
(\citealt{Mahmoud_2018,Mahmoud_2018b}). Another issue is that the observed power spectra often span a very broad range in frequencies, which is difficult to quantitatively match by the fairly small range of radii spanned by the hot flow without going to extreme parameters (\citealt{Ingram_2011,Ingram_2012,Mahmoud_2018, Mahmoud_2018b}).

A key to matching both the power spectral shape and width was the recognition that the disc generates considerable variability in the low/hard state, in addition to that expected from the hot flow (\citealt{Wilkinson_2009, Uttley_2011}). 
While the disc does not contribute to the {\it RXTE} bandpass ($>3$~keV) in the low/hard state, its variability will propagate down into the hot flow, so it will strongly affect the variability properties. 
\cite{Rapisarda_2016} proposed that the inner edge of the disc had a much longer variability time-scale than the outer edge of the hot flow due to its smaller scale height, and showed that this naturally produces double-peaked power spectra (but see \citealt{Veledina_2016} for another potential mechanism).
The slowly variable disc also widens the range of time-scales on which the X-ray flux varies even when the truncation radius is only a few tens of gravitational radii (\citealt{Rapisarda_2016}).

Modelling of the broad-band X-ray variability demonstrates how the timing properties give additional information about the nature of the accretion flow. Combining these with spectra (spectral-timing studies) gives an even more powerful tool, as it uses all the information from the 
energy spectrum and its fluctuations (power spectra) together with causal connections (lags/leads e.g. \citealt{Axelsson_2018, Mahmoud_2019, Wang_2021, DeMarco_2021}). In our previous work \cite[hereafter K22]{Kawamura_2022}, we developed a spectral-timing model based on propagating fluctuations from a turbulent disc through a spectrally-inhomogeneous (approximated by two Comptonisation regions) flow which generates variability at each radius.
We also incorporated reverberation of the variable Comptonisation components illuminating the disc to perform a self-consistent spectral-timing analysis.
We applied the model to the recently discovered black hole transient MAXI~J1820+070 (\citealt{Kawamuro_2018, Tucker_2018}), which has been widely studied (e.g. \citealt{Kara_2019, Shidatsu_2019, Homan_2020, Bright_2020, Axelsson_2021, Ma_2021, You_2021, Tetarenko_2021, Wang_2021, Prabhakar_2022}) thanks to its exceptional brightness, low galactic absorption (\citealt{Uttley_2018}), and intensive monitoring by multiple telescopes.
K22 fit the time-averaged energy spectrum for the {\it NICER} (0.5--10~keV) + {\it NuSTAR} (3--73~keV) and used this to develop a model for the variability below 10~keV seen in {\it NICER}. However, {\it NuSTAR} has less capability for fast timing, so K22 could not investigate 
the variability at higher energies, which means that we could not fully probe the innermost parts of the hot flow.
Better constraints on propagation require extending the bandpass for fast timing to higher energies.

Here we use contemporaneous data from {\it Insight-HXMT} (Section~\ref{sec:data}) to test our model at higher energies. 
We predict the high-energy power spectra and phase lags and show how these fail to  describe several key features of the data (Section~\ref{sec:old}). 
We give the model maximal freedom by fitting only the timing data rather than using the full spectral-timing data, and consider several ways to extend our propagating fluctuation model to better match the data (Section~\ref{sec:rev}). 
In particular, the phase lags give a clear indication that the propagation time through the flow is slower than the time-scale on which fluctuations are generated (\citealt{Rapisarda_2017a}), but the full energy dependence of the variability is quite difficult to fit. 
The key to matching the power spectra is to allow the Comptonisation spectra to pivot, so that they change in shape as well as normalisation in response to the fluctuations (\citealt{Mastroserio_2018, Mastroserio_2019, Mastroserio_2021}). 
This is physically expected from Comptonisation models (\citealt{Veledina_2016, Veledina_2018}) and is observed (\citealt{Malzac_2003, Gandhi_2008, Yamada_2013, Bhargava_2022}). 
We are able to get a good match to the timing properties (power spectra and phase-lag spectra) from 2.6~keV up to 48~keV by including spectral pivoting, as well as separation of generator and propagation time-scale.  
We implement this as a full spectral-timing model and find we can fit all the data in the 2.6--48~keV bandpass, though the model for both spectra and timing diverge from the data above this energy (Section~\ref{sec:spec_timing}).
We discuss the physical properties of the accretion flow, comparing them with theoretical hot flow models (Section~\ref{sec:discussion}), and then conclude that the current data quality is still better than the best physical models of the flow, which motivates further development (Section~\ref{sec:conclusions}). 
All of the technical details of the model formalism are given in the appendices so that the main text stresses the physical aspects of the model.


\section{Observation and data reduction}
\label{sec:data}

\begin{figure}
	\includegraphics[width=1.\columnwidth]{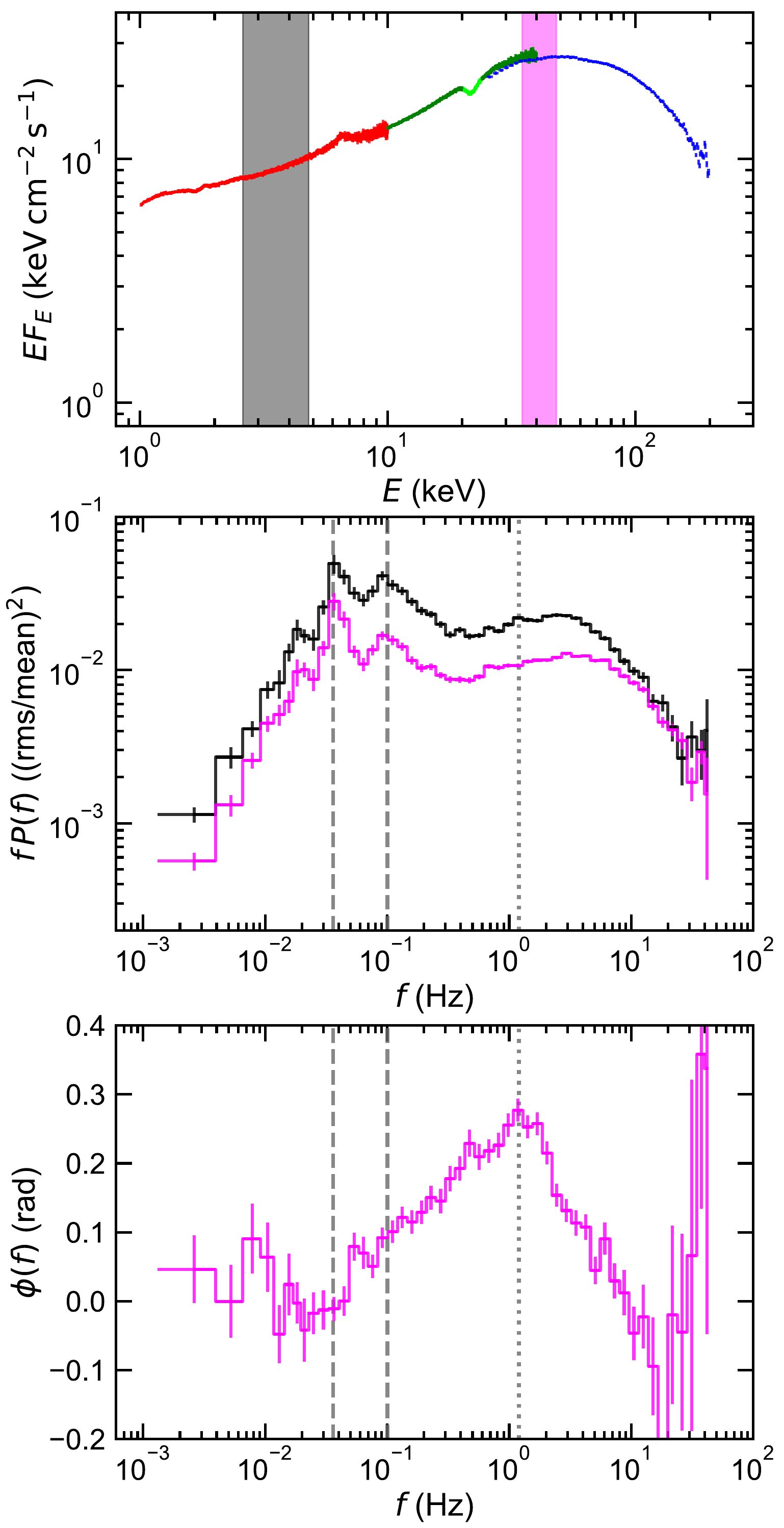}
	\caption{Spectral-timing properties of MAXI~J1820+070 observed by {\it Insight-HXMT}.
	{\it Top}: Time-averaged energy spectrum.
	Red, green, and blue markers represent LE, ME, and HE telescopes, respectively.
	The dip around 22~keV (light green) is associated with fluorescent lines of silver generated within the ME detector.
	The coloured regions show the low (black, 2.6--4.8~keV) and high (magenta: 35--48~keV) used to extract light curves.
	{\it Middle}: Power spectra calculated for low and high energy bands. 
        Both these have the characteristic double peak shape, but with the QPO and its harmonic (marked with dashed lines) superimposed. 
        The high-energy power spectrum is very similar in shape to that at low energy, but with lower normalisation (compare to the model in Fig.~\ref{fig:propagating_fluctuations_schematic} (f)).
	{\it Bottom}: Phase-lag spectrum between the light curves in the low and high energy bands.
	The lags are defined as positive if variations in higher energy bands lag behind those in lower energy bands (hard lags).
	The frequency at which the phase lag is maximum is marked with a dotted line. This is substantially lower than the characteristic frequency of the second peak in the power spectrum (compare to the model in Fig.~\ref{fig:propagating_fluctuations_schematic} (h)).
	}
	\label{fig:observation_data}
\end{figure}

We investigate the bright low/hard state of MAXI J1820+070 observed by {\it Insight-HXMT}.
2018-03-22 10:46:53 to 2018-03-24 02:49:49 (obsID P0114661003).
The same data are studied in \cite{Wang_2020, Ma_2021, Yang_2022}.
The observation time is slightly later than that we studied in K22 (Obs. ID: 1200120106; 2018-03-21), but there are simultaneous {\it NICER} data (Obs. ID: 1200120108; 2018-03-23) corresponding to these {\it Insight-HXMT} data. 
We checked that the energy spectrum, power spectra, and phase-lag spectra in these simultaneous 
{\it NICER} data are almost identical to K22.

We used the {\it Insight-HXMT} Data Analysis Software package (HXMTDAS) v2.04 to 
calibrate and screen the data using the same criteria as in \cite{Yang_2022}. 
We checked that the spectral and variability properties did not change substantially over the 
observation and then merged all the data together to achieve high signal-to-noise.

The resulting energy spectrum is shown in Fig.~\ref{fig:observation_data} (top).
The different colours represent different telescopes (red: LE, green: ME, blue: HE; \citealt{Zhang_2020}).
The energy spectrum from ME has a dip around 22~keV (light green), which is associated with silver fluorescent lines generated within the detector (\citealt{Li_2018}). Following \cite{You_2021}, we added 1.5~\% systematic errors to all spectral data.

To study the fast variability, we split the background-subtracted light curves into segments of $256\,\si{s}$ with $1/128\,\si{s}$ time bins ($2^{15}$ points), where we avoided any data gaps.
We only used the data where all telescopes were active to calculate light curves, 
using the same time selection for every energy band. We calculate the white-noise-subtracted power spectra and the cross spectra 
from each $256$~s segment and average them over different segments and logarithmically-spaced Fourier frequencies (\citealt{Uttley_2014, Ingram_2019}). All power spectra are normalised such that their integral over frequency corresponds to the fractional variance (\citealt{Miyamoto_1991, Vaughan_2003}).
Phase-lag spectra are calculated from the cross spectra, using the relation between the phase-lag spectrum $\phi (f)$ and cross spectrum $C(f)$, $\phi (f)=\tan ^{-1} (\Im[C(f)]/\Re[C(f)])$, where $\Re[\cdots]$ and $\Im[\cdots]$ denote the real and imaginary parts, respectively. Phase-lag relates to time lag via the relation $2\pi f t(f)=\phi(f)$.

Fig.~\ref{fig:observation_data} (middle) shows the power spectra of the 2.6--4.8~keV (black) and 35--48~keV (magenta) light curves
These energy bands are marked in the energy spectrum with shaded regions.
A QPO and its harmonic exist around 0.036~Hz and 0.1~Hz 
(shown with dashed lines), in addition to the broad-band variability.

Fig.~\ref{fig:observation_data} (bottom) shows the phase-lag spectrum between these two energy bands. 
The convention throughout this paper is that positive lags mean that the harder energy band lags behind the softer one.
The phase lag peaks 
at $\sim 1.2\,\si{Hz}$ (shown with a dotted line), which is not at the same frequency as the high energy peak in the power spectra.
This is unexpected as simple propagating fluctuations models have the same peak frequency both in the power spectrum and cross spectrum (\citealt{Ingram_2013, Rapisarda_2016}).
The QPO fundamental 
appears to affect the phase lag between these two energy bands, creating a dip in the phase-lag spectrum around the corresponding frequency (\citealt{Ma_2021}).
The effect of the second harmonic on the phase lag is not so clear between these energy bands, but we note it does have an impact for different choices of energy bands (\citealt{Ma_2021}).

For all of data fits performed in this paper, we use {\tt XSPEC} 12.12.1 (\citealt{Arnaud_1996}).
We formatted variability data and created a diagonal dummy response such that {\tt XSPEC} can import power spectra and phase-lag spectra as a function of Fourier frequency.
We developed our model as an {\tt XSPEC} model.
Being able to perform timing fits with the common tool in spectral fits is beneficial in performing spectral-timing fits.
For example, we will perform a joint fit to energy spectrum, six power spectra and five phase-lag spectra in Section~\ref{sec:spec_timing}.
We ignore variability below $\sim 10^{-2}\,\si{Hz}$ because it behaves differently from other Fourier frequencies.
\cite{Yang_2022} interpreted this low-frequency variability as the QPO sub-harmonic.


\begin{figure*}
	\includegraphics[width=\linewidth]{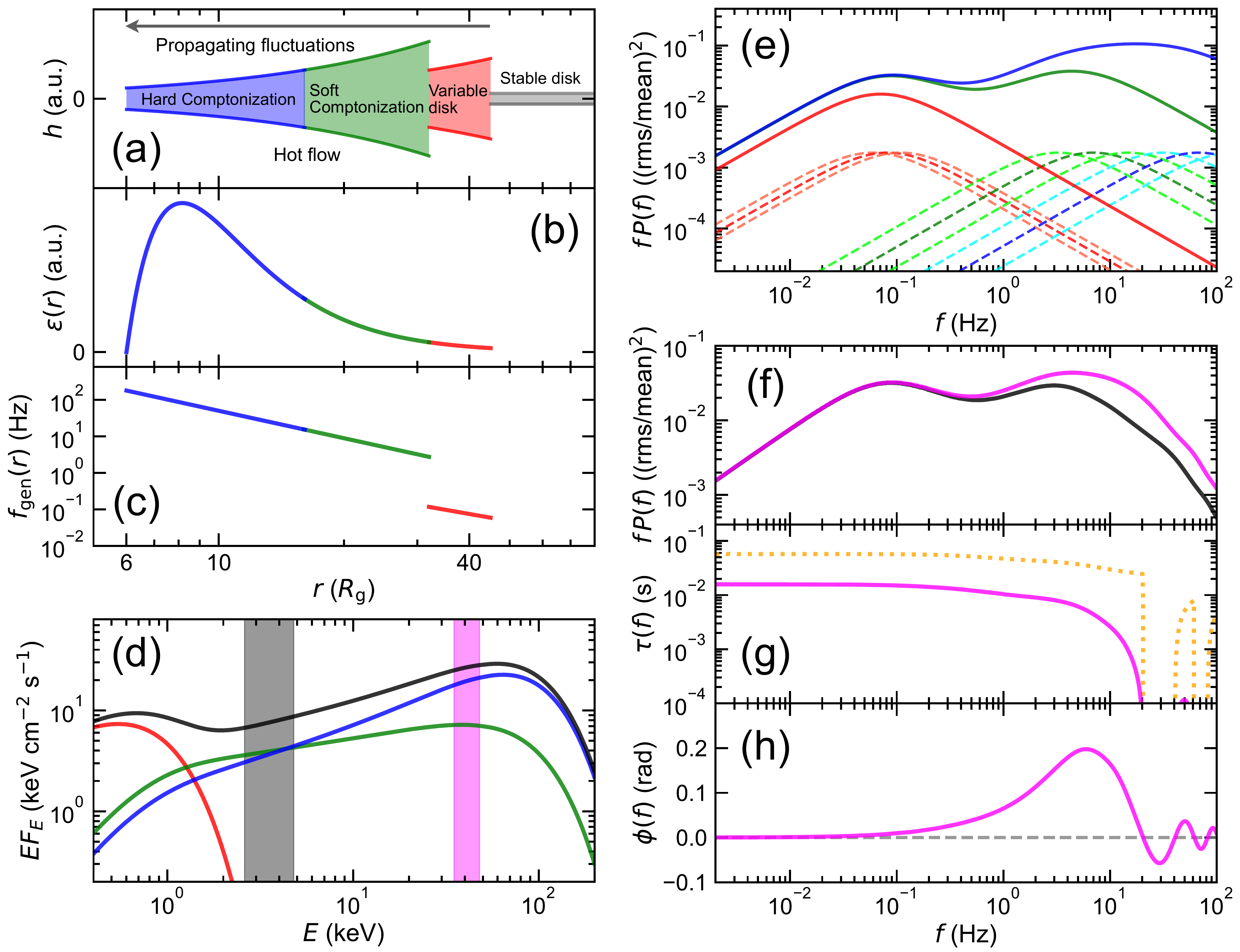}
	\caption{Model setup and predictions from K22.
	Here we show only the intrinsic components, rather than including reflection/reverberation, so as to focus on the physics of the propagation. 
    (a) Assumed accretion flow geometry (height as a function of radius). 
    There is an outer stable disc (grey) which is highly turbulent on its inner edge, forming the variable disc region. Inwards of this is the turbulent hot flow. 
    (b) Radial emissivity, assumed to be similar to that of a thin disc. 
    (c) Frequency at which variability is generated at each radius. 
    There is a discontinuity between the variable disc and hot flow as their scale heights are different, so their characteristic time-scales should be different. 
    Sample parameter values are $(B_{\mathrm{f}}, m_{\mathrm{f}})=(4, 1)$ for the hot flow and $(B_{\mathrm{d}}, m_{\mathrm{d}})=(0.03, 0.5)$ for the variable disc. 
    (d) Time-averaged energy spectra modelled (black) assuming a disc for the stable and variable disc emission (red), while the hot flow is assumed to be approximated by two Comptonisation components, soft (green) and hard (blue). 
    The relative luminosity in each component, together with the emissivity in (b), roughly sets the size scale of each region, so that the hard Comptonisation is for $r_{\mathrm{in}}\textrm{--}r_{\mathrm{sh}}=6\textrm{--}16$, the soft Comptonisation for $r_{\mathrm{sh}}\textrm{--}r_{\mathrm{ds}}=16\textrm{--}32$ and the variable disc for $r_{\mathrm{ds}}\textrm{--}r_{\mathrm{out}}=32\textrm{--}45$.
    (e) Sample power spectra of the local mass accretion rate in each spectral region. 
    The dashed lines represent the variability generated at each radius, with $r=45, 38, 33$ (in the variable disc: red), $30, 22, 16$ (in the soft Comptonisation: green), $12, 9, 6$ (in the hard Comptonisation: blue) from left to right.
    The frequency, at which each radius $f P(f)$ has its peak, corresponds to the local generator frequency $f_{\mathrm{gen}}(r)$.
	The solid lines show the total (generated plus propagated) variability at $r=38$ (red), $22$ (green), and $9$ (blue). 
	(e) Power spectra for two energy bands highlighted in (d).
	Any given energy band is not just a single component. 
	The low energy band (black) contains roughly equal amounts of soft and hard Comptonisation, while the high energy band (magenta) has mostly hard Comptonisation, but with some contribution from the soft Comptonisation as well. 
	Thus the power spectra of the light curves in the low and high energy bands are more similar than those of the soft and hard Comptonisation components in (e). 
	Nonetheless, there is still more high-frequency power in the high energy band than in the low energy band, but at lower frequencies the power spectra are identical as both contain the same propagated power. (g) Time-lag spectra for the low and high energy band light curves. 
	The orange dotted line shows the intrinsic lag of the soft Comptonisation light curve compared to the hard Comptonisation light curve. 
	This is $\sim 50$~ms, which is longer than the measured lag of the high energy band light curve versus the low energy band due to the mixture of spectral components in each band. 
	(h) Lag between high and low energy bands as a phase lag rather than a time lag (related by $\phi(f)=2\pi f\tau(f)$). 
	This peaks around the frequency of the high energy peak in the power spectrum.
    }
    \vspace{20pt}
	\label{fig:propagating_fluctuations_schematic}
\end{figure*}

\begin{table*}
\caption{Summary of our model parameters.
The variable flow is spectrally composed of the variable disc region and soft (outer) and hard (inner) Comptonisation regions.
We call the entire Comptonisation region the hot flow.
}
\begin{tabular}{llll}
\hline
Symbol                                 &Meaning                                              &Units            &Default                     \\
\hline
$M_{\mathrm{BH}}$                                                 &Black hole mass.                                                                      &$M_{\odot}$      &$8$                         \\
$N_{\mathrm{r}}$                                                  &Number of rings splitting the variable flow.                                           &                 &$40$                        \\
$r_{\mathrm{in}}$                                                 &Inner radius of the hard Comptonisation.                                              &$R_{\mathrm{g}}$ &$6$                         \\ 
$r_{\mathrm{sh}}$                                                 &Transition radius between the hard Comptonisation and soft Comptonisation.            &$R_{\mathrm{g}}$ &$16$                        \\
$r_{\mathrm{ds}}$                                                 &Transition radius between the disc and soft Comptonisation.                           &$R_{\mathrm{g}}$ &$32$                        \\
$r_{\mathrm{out}}$                                                &Outer radius of the variable disc.                                                   &$R_{\mathrm{g}}$ &$45$                        \\
$F_{\mathrm{var, f}}$ ($F_{\mathrm{var, d}}$)                     &Fractional intrinsic variability per radial decade in the hot flow (variable disc).  &                 &$0.8$                         \\
$D$                                                               &Damping factor.                                                                       &                 &$0$                           \\
$B_{\mathrm{f}}$ ($B_{\mathrm{d}}$)                               &Coefficient of the generator frequency in the hot flow (variable disc).                &                 &$0.03$                        \\
$m_{\mathrm{f}}$ ($m_{\mathrm{d}}$)                               &Power-law index of the generator frequency in the hot flow (variable disc).            &                 &$0.5$                         \\
$B^{(\mathrm{p})}_{\mathrm{f}}$ ($B^{(\mathrm{p})}_{\mathrm{d}}$) &Coefficient of the propagation frequency in the hot flow (variable disc).            &                 &$0.03$                        \\
$m^{(\mathrm{p})}_{\mathrm{f}}$ ($m^{(\mathrm{p})}_{\mathrm{d}}$) &Power-law index of the propagation frequency in the hot flow (variable disc).        &                 &$0.5$                         \\
$\gamma$                                                          &Power-law index of the emissivity.                                                    &                 &3                           \\
$b(r)$                                                            &Inner boundary condition of the emissivity.                                           &                 &$1-\sqrt{r_{\mathrm{in}}/r}$\\
$t_{0, \mathrm{h}}$ ($t_{0, \mathrm{s}}$)                         &Time delay of the top hat impulse response of reverberation for the hard (soft) Comptonisation$^a$.    &$\si{s}$&$5.5\times 10^{-3}$\\
$\Delta t_{0, \mathrm{h}}$ ($\Delta t_{0, \mathrm{s}}$)           &Time duration of the top hat impulse response of reverberation for the hard (soft) Comptonisation$^a$. &$\si{s}$&$10 \times 10^{-3}$\\
$S_{0}(E)$                                                        &Fractional contribution of spectral components to the flux$^{b,c,d}$.                 &                 &0.5                           \\
$\eta _{\mathrm{0, h}}$ ($\eta _{\mathrm{0, s}}$)                 &Constant term of the sensitivity of the hard (soft) Comptonisation to change in mass accretion rate$^e$.                &                 &1                            \\
$\eta _{\mathrm{1, h}}$ ($\eta _{\mathrm{1, s}}$)                 &Gradient term of the sensitivity of the hard (soft) Comptonisation to change in mass accretion rate$^e$.                &                 &0                            \\
\hline
\multicolumn{4}{l}{{\it Notes.} $^a$ Parameters are required when the reverberation is considered.}\\
\multicolumn{4}{l}{$^b$ Each spectral component has its own parameter: $S_{0}(E)$ consists of $S_{\mathrm{d}}(E)$, $S_{\mathrm{s}}(E)$, $S^{(\mathrm{r})}_{\mathrm{s}}(E)$, $S_{\mathrm{h}}(E)$, and $S^{(\mathrm{r})}_{\mathrm{h}}(E)$ (see Section~\ref{sec:old}).}\\
\multicolumn{4}{l}{$^c$ $S_{0}(E)$ is replaced by $\eta (E) S_{0}(E)$ for timing fits when the spectral pivoting is included (Section~\ref{sec:rev2}).}\\
\multicolumn{4}{l}{$^d$ $S_{0}(E)$ is calculated from spectral models for spectral-timing fits (Section~\ref{sec:spec_timing}).}\\
\multicolumn{4}{l}{$^e$ Parameters are required for spectral-timing fits (Section~\ref{sec:spec_timing}).}\\
\end{tabular}
\\
\label{tab:model_parameters}
\end{table*}

\section{Propagation and reverberation in our previous model}
\label{sec:old}

\subsection{Summary of our previous work}

We start with a summary of our previous physical model. Fundamentally 
we assume that variability is generated by fluctuations in density in the flow, which propagate inwards as accretion rate fluctuations. Thus the variability generated at each radius propagates down with the accretion flow so that slower fluctuations generated at large radii imprint on faster fluctuations generated at small radii. These fluctuations in mass accretion rate change the luminosity emitted in the spectrum at that radius. 

We then need additional assumptions to turn this into a quantitative model, as we have to assume the form of radial stratification for both the spectrum and variability, and how fluctuations are generated and propagated. We expand on each of these below.

We assume a basic geometry which is a truncated disc/hot inner flow, as shown in 
Fig.~\ref{fig:propagating_fluctuations_schematic} (a) with emissivity at each radius set by the Shakura-Sunyaev thin disc approximation (Fig.~\ref{fig:propagating_fluctuations_schematic} (b)). 
These two assumptions alone are enough to roughly set the transition radius from the inner edge of the disc to the outer edge of the hot flow by energetics to $\sim 45R_{\mathrm{g}}$, though actually instead we set this from the QPO frequency, assuming Lense-Thirring precession (\citealt{Ingram_2009}). 
The similarity of the two estimates gives support to the Lense-Thirring interpretation. 
In all the following, we use the convention that $R=rR_{\mathrm{g}}$, where $R_{\mathrm{g}}=GM_{\mathrm{BH}}/c^2$.

Spectrally, we assume that the flow emits a single component at each radius. This might be unique to each radius, with e.g. each radius in the disc emitting a blackbody with temperature $T(R)$, while the hot flow emits a Comptonised spectrum whose parameters (electron temperature, optical depth) scale smoothly with radius. However, spectral models are quite degenerate so we approximate the emission from the truncated disc region 
($r_{\mathrm{out}}$--$r_{\mathrm{ds}}$) as
a disc blackbody, and we 
approximate the hot flow as two zones as 
physically we do expect that there are two main regions in the flow. Close to the disc, seed photons for Comptonisation are predominantly from the disc. However, it is quite easy for this Comptonisation to become optically thick along the equatorial direction, shielding the inner regions from the disc photons so that seed photons are predominantly from cyclo-synchrotron (\citealt{Poutanen_2014}). Thus we assume that radii from 
$r_{\mathrm{ds}}$--$r_{\mathrm{sh}}$ emit soft Comptonisation, while radii from  $r_{\mathrm{sh}}$--$r_{\mathrm{in}}$ emit hard Comptonisation ($r_{\mathrm{in}}<r_{\mathrm{sh}}<r_{\mathrm{ds}}<r_{\mathrm{out}}$; Fig.~\ref{fig:propagating_fluctuations_schematic} (a), (d)). 
Both Comptonisation components illuminate the disc to produce reflection, while the energy not reflected is (mostly) thermalised, enhancing the cool disk emission. 
K22 show that these assumptions give a good fit to the energy spectrum from $0.5\textrm{--}80$~keV.

The variability is also assumed to be radially stratified such that each radius generates fluctuations with a characteristic frequency, $f_{\rm gen}(r)$. 
This is assumed to have a power-law form as a function of radius in the hot flow, 
so $f_{\mathrm{gen}}(r)=Br^{-m}f_{\mathrm{K}}(r)$, where $f_{\mathrm{K}}(r)=(1/2\pi) r^{-3/2}$ in units of $c/R_{\mathrm{g}}$. 
The power-law scaling parameters are allowed to be different 
between the hot flow ($B_{\mathrm{f}}, m_{\mathrm{f}}$) and disc ($B_{\mathrm{d}}, m_{\mathrm{d}}$), giving a different time-scale to reflect the different scale heights of the two flows (\citealt{Rapisarda_2016}).
Thus, the generator frequency is modelled with 
\begin{equation}
    f_{\mathrm{gen}} (r)=
    \begin{cases}
        B_{\mathrm{f}} r ^{-m_{\mathrm{f}}} f_{\mathrm{K}} (r) & (r_{\mathrm{in}} \leq r <r_{\mathrm{ds}})  ,\\
        B_{\mathrm{d}} r ^{-m_{\mathrm{d}}} f_{\mathrm{K}} (r) & (r_{\mathrm{ds}} \leq r < r_{\mathrm{out}}),
    \end{cases}
    \label{eq:viscous_frequency}
\end{equation}
as shown in Fig.~\ref{fig:propagating_fluctuations_schematic} (c). 
We assume each logarithmic radial interval generates the same amplitude of variability. 
The fluctuations generated at each radius propagate down without losses
to produce a fluctuating mass accretion rate at each radius which modulates the emitted luminosity.

As noted above, we assume the QPO is set by Lense-Thirring precession of the entire hot flow, and the first bump in the power spectrum is set by the turbulent disc (\citealt{Ingram_2011}).
This sets $(B_{\mathrm{d}}, m_{\mathrm{d}})=(0.03, 0.5)$ and $r_{\mathrm{out}}=45$.

The dashed lines in Fig.~\ref{fig:propagating_fluctuations_schematic} (e) show three sample power spectra for the generated variability of the local mass accretion rate in each region (variable disc: red, soft Comptonisation: green, hard Comptonisation: blue, the middle ring of each highlighted in a darker colour). 
The functional form is a zero-centred Lorentzian with the cut-off frequency corresponding to the local generator frequency $f_{\mathrm{gen}}(r)$, which yields the peak at $f_{\mathrm{gen}}(r)$ in the $fP(f)$ representation. 

K22 assumed that the propagation time-scale was the same as that on which the fluctuations were generated and
called this the viscous time-scale (\citealt{Lyubarskii_1997, Arevalo_2006, Ingram_2009}). 
This assumption sets the propagation speed at any radius $v_{\mathrm{p}}(r) = r f_{\mathrm{gen}}(r)$ in units of $c$. 
However, here we will revisit this assumption, so to avoid confusion, we do not use the term `viscous time-scale' but use `generator time-scale' and `propagation time-scale' in this paper to make it clear which one we mean.
K22 also assumed that the fluctuating energy release only changed the normalisation of the spectral component emitted at that radius, not its shape. 

The solid lines in Fig.~\ref{fig:propagating_fluctuations_schematic} (e) show the propagated (total) power spectra from each region (disc: red, soft Comptonisation: green, hard Comptonisation: blue). 
This is not the same as the power spectrum in any given energy band as Fig.~\ref{fig:propagating_fluctuations_schematic} (d) shows that each energy band contains a mix of components.

Fig.~\ref{fig:propagating_fluctuations_schematic} (f) shows the power spectra for the two chosen energy bands, low (2.6--4.8 keV: black) and high  (35--48 keV: magenta), also highlighted in the same colours in  Fig.~\ref{fig:propagating_fluctuations_schematic} (d). 
While neither band contains the disc emission component, both bands contain the propagated disc variability. The low energy band also contains both soft and hard Comptonisation, so has the generated/propagated power in the soft Comptonisation region, plus some of the highest frequency power generated in the hard Comptonisation region. 
The high energy band is dominated by the hard Comptonisation emission, with variability which is propagated down from both the disc and soft Comptonisation region, plus the highest frequency power generated/propagated through the hard Comptonisation region. 
Thus the power spectra are almost identical for below $\sim 1\,\si{Hz}$, indicating that variability on these slow time-scales is propagated from the outer regions rather than generated at their emission regions, while they diverge at the highest frequencies where the low energy band does not include as much of the hard Comptonisation component as the high energy band. 

The propagation time-scale is explicitly seen in the time lag (Fig.~\ref{fig:propagating_fluctuations_schematic} (g)). 
The luminosity weighted mean radius of the soft Comptonisation band is 22~$R_{\mathrm{g}}$, whereas that for the hard is 11~$R_{\mathrm{g}}$. This lag time (integrating $1/(rf_{\mathrm{prop}}(r))$) is 51~ms, as seen as the value of the approximately constant time lag at low frequencies
(orange dashed line). 
The time lag drops when the variability starts to be dominated by the generated variability (which is not correlated) rather than the propagated variability i.e. at approximately the generator time-scale of the hard Comptonisation region. 
At the highest frequencies, the lag also produces oscillatory structure at high frequencies, by the interference generated by the mass accretion rate in the hard Comptonisation region being the same as in the soft Comptonisation region, but lagged by the mean propagation time-scale. 
This oscillatory structure has a period of $~50$~ms.
However, the intrinsic lag from the spectral components is diluted by a factor $\sim 3$ when looking at the low and high energy bands rather than the soft and hard Comptonisation components. 
This is because the low energy band includes both soft and hard Comptonisation, while the high energy band is dominated by the hard Comptonisation. 
While the value of the measured lag time changes, the oscillatory structure period remains at the propagation time-scale. 

This oscillation can be seen more clearly in Fig.~\ref{fig:propagating_fluctuations_schematic} (h), which shows the phase lag rather than time lag, so each time-scale is multiplied by the factor $2\pi f$.
In this representation, the phase lag peaks at a frequency between the high-frequency peak in the power spectra at low and high energy.
By comparison to the panel above, it is also clear that the oscillatory structure at high frequencies has the same period as in the undiluted (orange) time lags.

All model parameters are summarized in Table~\ref{tab:model_parameters} and
we give a more quantitative model summary in Appendix~\ref{sec:app_model_summary}.

The fluctuating soft and hard Comptonisation regions illuminate the outer disc and produce a reflected/reprocessed signal which lags behind the generated and propagated flow variability by the light travel time to the disc.
The reflected emission itself is not strong, though this real reverberation signal will add to the  soft lags produced by interference in the propagation (hard) lags seen in Fig.~\ref{fig:propagating_fluctuations_schematic} (g). However, at lower energies, the photons which are not reflected heat the disc, giving a thermal reverberation signal which is strong at energies close to that of the disc emission ($\lesssim 2\,\si{keV}$; \citealt{Kara_2019}).
This reverberation signal gives an independent check on the assumption of the disc truncation radius, and the fact that it is consistent (\citealt{DeMarco_2021}) gives strong supporting evidence for the underlying assumption that the QPO mechanism is Lense-Thirring precession.
Our previous model includes the reverberation, along with the propagating fluctuations (K22).

K22 showed that this model gave a fairly good fit to the energy dependence of the power spectrum across the {\it NICER} energy band (0.5--10~keV) and to the lags between the same fluctuations in different energy bands as a function of frequency. 
However, while the spectral components were built from {\it NuSTAR} data, which extended above 10~keV, this instrument does not have a sufficient area to do high-time-resolution studies, so the model prediction at higher energies could not be tested. 

This outburst of MAXI~J1820+070 was also monitored by {\it Insight-HXMT} (\citealt{Ma_2021, You_2021, Yang_2022}), which does have a sufficient effective area at high energies. 
As mentioned in the previous section, the {\it Insight-HXMT} data are not absolutely simultaneous with the {\it NICER}/{\it NuSTAR} dataset we used in K22, but they are very close in time, and spectral-timing properties are nearly constant during these periods.
Hence we take the spectral-timing model of K22, use it to predict the higher energy behaviour, and compare it to the {\it Insight-HXMT} data.


\begin{figure*}
	\includegraphics[width=\linewidth]{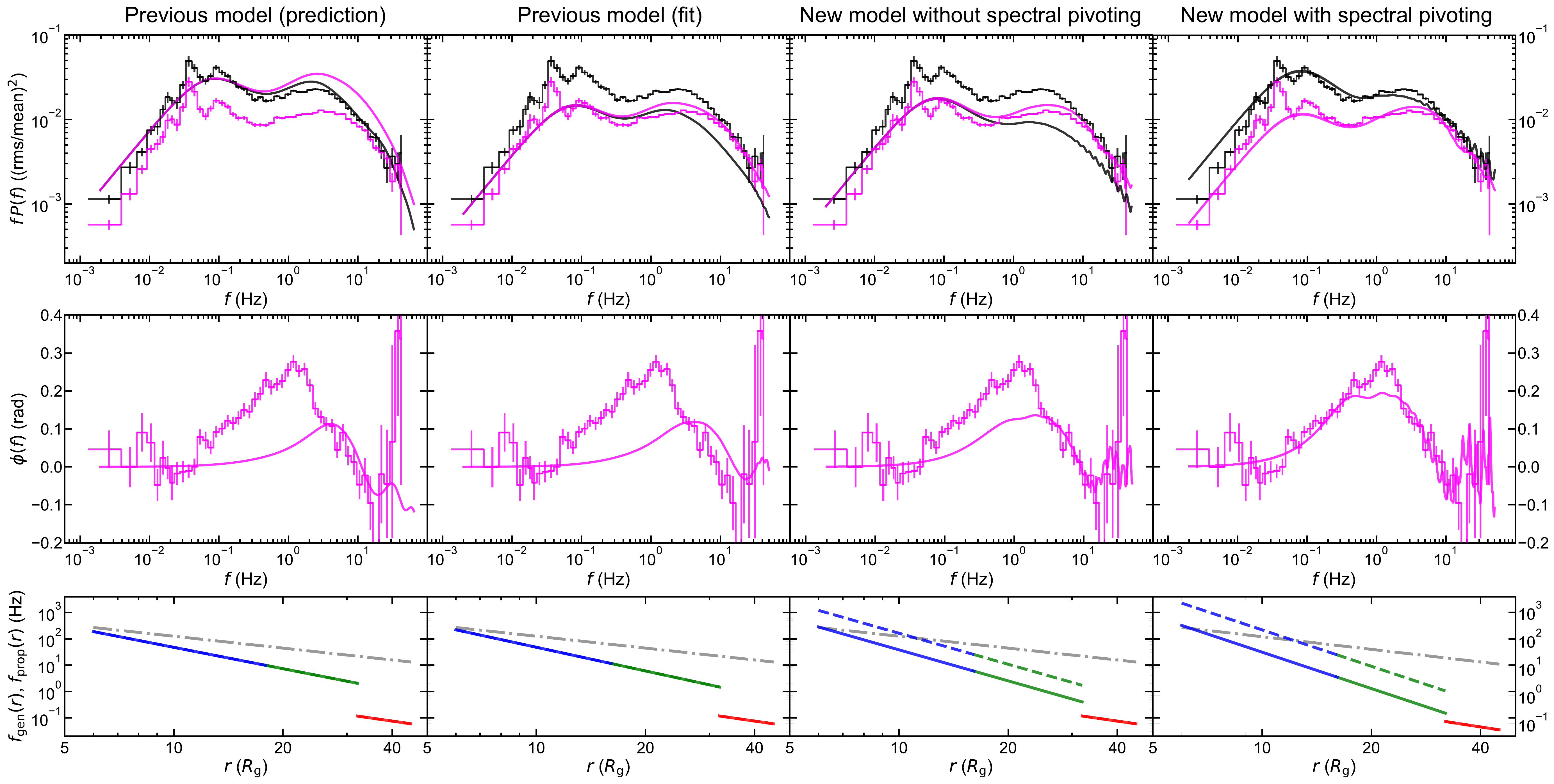}
	\caption{The effect of the model updates on the timing properties.
	Each figure shows the low (2.4--4.8~keV: black) and high (35--48~keV: magenta) energy band power spectra (upper) and phase lag spectra (middle), with data shown as the stepped line with errors and the model as the smooth curve. The lower panel shows the propagation frequency (solid) and generator frequency (dashed) used in the model calculations, with the Keplerian frequency (dash-dotted) for reference.
    {\it Left}: Predictions from the previous model from K22 (Section~\ref{sec:old}) built from a full spectral-timing fit to the 0.5--10~keV data.
    {\it Mid-left}: Fitting with the previous model (Section~\ref{sec:old}), ignoring the time averaged spectrum.
	{\it Mid-right}: Extending the model to include a different propagation and generator time-scale (\citealt{Rapisarda_2017a}). 
	This shifts the frequency of the phase lag peak but does not change the power spectra. 
    We also gave the model the freedom to include damping \citealt{Mahmoud_2018b}, but the best fit value was close to zero, so this is not shown (Section~\ref{sec:rev1}). 
    {\it Right}: Including spectral pivoting as well as a difference in generator and propagation time-scale (Section~\ref{sec:rev2}).
    This allows the power spectral normalisation of the high energy band to be lower than at low energies, giving a significant improvement in the consistency of the model calculations.}
	\label{fig:psd_phase_fit_comp}
\end{figure*}

\begin{table}
\caption{Model parameter values used in Figs.~\ref{fig:psd_phase_fit_comp}. 
Common parameter values used in all fitting are $M_{\mathrm{BH}}=8$, $N_{\mathrm{r}}=40$, $r_{\mathrm{in}}=6$, $r_{\mathrm{out}}=45$, $B_{\mathrm{d}}=B^{(\mathrm{p})}_{\mathrm{d}}=0.03$, $m_{\mathrm{d}}=m^{(\mathrm{p})}_{\mathrm{d}}=0.5$, $\gamma=3$, $b(r)=1-\sqrt{r_{\mathrm{in}}/r}$.
The transition radii are fixed to $r_{\mathrm{sh}}=17.8$, $r_{\mathrm{ds}}=32.1$ for the left column and $r_{\mathrm{sh}}=16$, $r_{\mathrm{ds}}=32$ for other columns, although these differences are too subtle to become important.
Other constraints are $F_{\mathrm{var, f}}=F_{\mathrm{var, d}}$ and $D=0$.
The mark, `(f)', means that a value of the corresponding parameter is fixed.}
\begin{tabular}{lllll}
\hline
Symbol&Left&Mid-left&Mid-right&Right$^{a}$\\
\hline
$F_{\mathrm{var, d}}$                                     &$0.8$ (f)           &$0.53$            &$0.59$       &$0.8$ (f)                                \\
$B_{\mathrm{f}}$                                          &$6$ (f)             &$11.9$            &$342$        &$560$                                    \\
$m_{\mathrm{f}}$                                          &$1.2$ (f)           &$1.50$            &$2.42$       &$2.73$                                   \\
$B^{(\mathrm{p})}_{\mathrm{f}}$                           &$=B_{\mathrm{f}}$   &$=B_{\mathrm{f}}$ &$80.1$       &$166$                                    \\
$S_\mathrm{d}(2.6\textrm{--}4.8\,\si{keV})$               &$0.001$ (f)         &$0$ (f)           &$0$ (f)      &$0$ (f)                                  \\
$S_\mathrm{s}(2.6\textrm{--}4.8\,\si{keV})$               &$0.356$ (f)         &$0.505$           &$0.471$      &$0.305$                                  \\
$S_\mathrm{h}(2.6\textrm{--}4.8\,\si{keV})$               &$0.330$ (f)         &$=1-S_{\mathrm{s}}(E)$&$=1-S_{\mathrm{s}}(E)$      &$0.571$                                  \\
$S^{(\mathrm{r})}_\mathrm{s}(2.6\textrm{--}4.8\,\si{keV})$&$0.307$ (f)         &$0$ (f)           &$0$ (f)      &$0$ (f)                                  \\
$S^{(\mathrm{r})}_\mathrm{h}(2.6\textrm{--}4.8\,\si{keV})$&$0.006$ (f)         &$0$ (f)           &$0$ (f)      &$0$ (f)                                  \\
$S_\mathrm{d}(35\textrm{--}48\,\si{keV})$                 &$0$ (f)             &$0$ (f)           &$0$ (f)      &$0$ (f)                                  \\
$S_\mathrm{s}(35\textrm{--}48\,\si{keV})$                 &$0.213$ (f)         &$0.348$           &$0.277$      &$-0.056$                                 \\
$S_\mathrm{h}(35\textrm{--}48\,\si{keV})$                 &$0.474$ (f)         &$=1-S_{\mathrm{s}}(E)$&$=1-S_{\mathrm{s}}(E)$      &$0.476$                                  \\
$S^{(\mathrm{r})}_\mathrm{s}(35\textrm{--}48\,\si{keV})$  &$0.134$ (f)         &$0$ (f)           &$0$ (f)      &$0$ (f)                                  \\
$S^{(\mathrm{r})}_\mathrm{h}(35\textrm{--}48\,\si{keV})$  &$0.179$ (f)         &$0$ (f)           &$0$ (f)      &$0$ (f)                                  \\
\hline
\multicolumn{5}{l}{{\it Notes.} $^a$ $S_{0}(E)$ means $\eta (E)S_{0}(E)$ in this column.}\\
\end{tabular}\\
\label{tab:model_parameter_values}
\end{table}

\subsection{Comparison of our previous model to {\it Insight-HXMT} data}

We compare the predictions of our previous model to the power spectra for the 2.6--4.8~keV and 35--48~keV bands and the phase-lag spectrum between these bands calculated from the {\it Insight-HXMT} observation data in Fig.~\ref{fig:psd_phase_fit_comp} (left).
Model parameter values are summarised in Table~\ref{tab:model_parameter_values}, which also contains those for the rest of the columns in Fig.~\ref{fig:psd_phase_fit_comp}.
The lower energy band is well reproduced by our previous model from K22, as expected, as it is within the {\it NICER} energy range over which K22 got good fits. 
However, the power spectrum at the higher energy band is clearly overestimated, and the phase lag between the two is completely wrong, peaking at too high a frequency with a lag which is too short to match the data. 

\section{Timing fits: exploring the additional processes required to match the high energy variability}
\label{sec:rev}

The specific spectral-timing model of K22 plainly cannot fit the data, so instead here we explore whether the model is capable of reproducing the observed timing properties. We give maximal flexibility by ignoring the time-averaged spectrum 
(but we will come back to joint spectral-timing modelling in Section~\ref{sec:spec_timing})
and attempt to minimize the sum of $\chi ^2$ values for the power spectra and phase-lag spectrum:
\begin{equation}
	\begin{split}
		\sum_{k} \biggl\{ &\left( \frac{P_{\mathrm{data}}(E_1, f_{k})-P_{\mathrm{model}}(E_1, f_{k})}{\Delta P_{\mathrm{data}}(E_1, f_{k})} \right) ^2 \\
		+ &\left( \frac{P_{\mathrm{data}}(E_2, f_{k})-P_{\mathrm{model}}(E_2, f_{k})}{\Delta P_{\mathrm{data}}(E_2, f_{k})} \right) ^2 \\ 
	+ &\left( \frac{\phi_{\mathrm{data}}(E_1, E_2, f_{k})-\phi_{\mathrm{model}}(E_1, E_2, f_{k})}{\Delta \phi_{\mathrm{data}}(E_1, E_2, f_{k})} \right) ^2 \biggr\}, \\ 
	\end{split}
\end{equation}
where $P_{\mathrm{data}}(E, f)$, $\Delta  P_{\mathrm{data}}(E, f)$ are the observed power spectrum and its one-sigma error at frequency $f$ for energy $E$, 
$\phi_{\mathrm{data}}(E, E', f)$, $\Delta  \phi_{\mathrm{data}}(E, E', f)$ the equivalents for the phase-lag spectrum between energy $E$ and $E'$,
$P_{\mathrm{model}}(E, f)$ and $\phi (E, E', f)$ the modelled power spectrum and phase-lag spectrum,
$E_{1}=2.6\textrm{--}4.8\,\si{keV}$, $E_{2}=35\textrm{--}48\,\si{keV}$, and $f_{k}$ the sampled Fourier frequency.

Our model requires the fraction of each spectral component to calculate the power spectra and cross spectra.
We express this fraction as $S_{\mathrm{d}}(E)$, $S_{\mathrm{s}}(E)$, $S^{(\mathrm{r})}_{\mathrm{s}}(E)$, $S_{\mathrm{h}}(E)$, $S^{(\mathrm{r})}_{\mathrm{h}}(E)$ for the variable disc, soft Comptonisation and its reflection, hard Comptonisation and its reflection, respectively.
Since X-ray energy spectra in the hard state are almost fully occupied by these five components at most (\citealt{Zdziarski_2021, Zdziarski_2021b, Zdziarski_2022}), we require the sum of these fractions to correspond to unity:
\begin{equation}
    S_{\mathrm{d}}(E)+S_{\mathrm{s}}(E)+S^{(\mathrm{r})}_{\mathrm{s}}(E)+S_{\mathrm{h}}(E)+S^{(\mathrm{r})}_{\mathrm{h}}(E)=1.
    \label{eq:spec_constraint}
\end{equation}  
Whereas in Fig.~\ref{fig:psd_phase_fit_comp} (left), these fractions were calculated from the result of spectral fit in K22 for the self-consistent spectral-timing modelling, here we let them be independent of the time-averaged spectrum in order to focus on the variability properties.
We fix $S_{\mathrm{d}}(E)=0$ because the disc emission is negligible above 2.6~keV.
We also ignore the reverberation, i.e., $S^{(\mathrm{r})}_{\mathrm{s}}(E)=S^{(\mathrm{r})}_{\mathrm{h}}(E)=0$, to simplify the model.
This is not a bad approximation for our purpose, as we want to capture the broad-band power spectra and hard lags from propagation. Reverberation makes only small changes to the variability properties on the energy and variability range of interest here. 

Finally, we only have the soft and hard Comptonisation components with the constraints of $S_{\mathrm{s}}(E)+S_{\mathrm{h}}(E)=1$.
We do not include any models for the QPO features for simplicity.

We keep the black hole mass of $M_{\mathrm{BH}}=8M_{\odot}$ (\citealt{Torres_2020}) and emissivity profile, i.e., $\gamma=3$ and $b(r)=1-\sqrt{r_{\mathrm{in}}/r}$ (\citealt{Shakura_1973, Novikov_1973}), where we assume that radiation energy from the annulus ranging from $r$ to $r+\Delta r$ is proportional to $r^{-\gamma}b(r)2\pi r \Delta r$.
In K22, the transition radii $r_{\mathrm{sh}}, r_{\mathrm{ds}}$ were calculated from the emissivity profile and spectral decomposition.
However, we lack spectral decomposition.
In addition, it turned out that model calculations are hardly affected by small changes in these parameters.
Thus, we simply fix these transition radii to typical values, $r_{\mathrm{sh}}=16$ and $r_{\mathrm{sh}}=32$.

We show the result of the joint fit to the power spectra for 2.6--4.8~keV and 35--48~keV and the phase-lag spectrum between these energy bands in Fig.~\ref{fig:psd_phase_fit_comp} (mid-left).
The fit is not qualitatively improved even giving the K22 model maximal freedom to fit without constraints from  the time-averaged energy spectrum.
The K22 model always has a high-energy-band power spectrum similar to that in the low energy band
everywhere except at the highest frequencies. 
Yet the data have very different power spectral normalisations even at low frequencies where propagation should dominate.

Plainly, while the previous model from K22 was designed to fit the data below 10~keV, it does not extrapolate to the higher energies, so does not adequately describe the physics of the propagation of fluctuations through the flow. 
This is important as K22 shows that the propagation speed is a key determinant of the nature of the hot flow, which can allow large-scale magnetically dominated flows (MAD) to be distinguished from those with turbulent dynamo (SANE) models. 
The poor applicability of our previous model to higher energy bands motivates our study to improve it.


\subsection{Suppressing variability at high energies with a constant spectral shape}
\label{sec:rev1}

The major feature missing in the previous model for the power spectra is the strong suppression of fractional variability at high energies. 
The generation/propagation of fluctuations in the model, where slower fluctuations generated outer regions propagate down through the flow, always leads to an increase in variability with energy, as long as the spectrum hardens inwards. 
In contrast, the {\it Insight-HXMT} observation data show that plainly the high-energy broad-band power spectrum is a factor $\sim 3$ lower than the low-energy power spectrum at all frequencies (Fig.~\ref{fig:observation_data} (middle)).
This decrease in fractional variability with energy was not seen in the {\it NICER} energy band ($\lesssim 10\,\si{keV}$; K22). 
But it has been seen before, in e.g. the {\it RXTE} data of other black hole binary low/hard states (e.g. \citealt{Nowak_1999, Axelsson_2018} for Cyg~X-1; \citealt{Malzac_2003} for XTE~J1118+480).
In the context of other propagating fluctuations models, it was modelled by the damping of high-frequency fluctuations as they propagate inwards (\citealt{Arevalo_2006, Rapisarda_2017a}), and by decreasing the intrinsic variability power generated in the inner regions (\citealt{Mahmoud_2018b}). 
To implement these effects in our model, we introduce two new parameters.
One is a damping parameter $D$, which suppresses high-frequency variability by $\mathrm{exp}(-D f \Delta t)$, where $\Delta t$ is the propagation time.
The damping effect is ignored if $D=0$.
We also allow the intrinsic variability amplitude to be different between the hot flow $F_{\mathrm{var, h}}$ and disc $F_{\mathrm{var, d}}$ (the previous model from K22 has $F_{\mathrm{var,f}}=F_{\mathrm{var,d}}$). 

Another observational feature that our previous model fails to capture is the discrepancy in the frequency at which the power spectra (in the $fP(f)$ representation) and phase-lag spectra peak.
The power spectra and phase-lag spectra calculated by the K22 model have a similar peak frequency.
This observational property is also seen in the {\it RXTE} data (e.g. XTE J1550-564: \citealt{Rapisarda_2017a}), where the proposed solution was to allow the propagation time-scale to be different to the generator time-scale on which the fluctuations are generated.
Following this, we separate these time-scales and define the propagation frequency with
\begin{equation}
    f_{\mathrm{prop}} (r)=
    \begin{cases}
        B^{(\mathrm{p})}_{\mathrm{f}} r ^{-m^{(\mathrm{p})}_{\mathrm{f}}} f_{\mathrm{K}} (r) & (r_{\mathrm{in}} \leq r_{n} <r_{\mathrm{ds}})  ,\\
        B^{(\mathrm{p})}_{\mathrm{d}} r ^{-m^{(\mathrm{p})}_{\mathrm{d}}} f_{\mathrm{K}} (r) & (r_{\mathrm{ds}} \leq r_{n} < r_{\mathrm{out}}),
    \end{cases}
    \label{eq:propagation_frequency}
\end{equation}
such that the propagation speed is provided by $v_{\mathrm{p}}(r)=rf_{\mathrm{prop}}(r)$.
It is difficult to constrain disc parameters as the disc emission has negligible contributions to the energy range of interest.
Thus we keep employing $f_{\mathrm{gen}}(r)=f_{\mathrm{prop}}(r)$ in the variable disc region with $(B^{(\mathrm{p})}_{\mathrm{d}}, m^{(\mathrm{p})} _{\mathrm{d}})=(B_{\mathrm{d}}, m_{\mathrm{d}})=(0.03, 0.5)$.
To reduce the number of free parameters, we assume that $f_{\mathrm{prop}}(r)$ has the same radial dependence as $f_{\mathrm{gen}}(r)$, i.e., $m^{(\mathrm{p})}_{\mathrm{f}}=m_{\mathrm{f}}$. 
Eventually, we have only one additional parameter $B^{(\mathrm{p})}_{\mathrm{f}}$.

The modified model formalism due to the damping effect is given in Appendix~\ref{sec:app_damping}.
Other additional effects, $F_{\mathrm{var, d}} \neq F_{\mathrm{var, f}}$ and $f_{\mathrm{gen}} (r) \neq f_{\mathrm{prop}}(r)$, just alter the power spectrum of intrinsic mass accretion rate variability at $n$th ring $|A(r_n, f)|^2$ (\ref{eq:mdot_power_intr}) and the propagation time from the outer $k$th ring to the inner $n$th ring $\Delta t_{k, n}$ (\ref{eq:prop_time}), respectively, without affecting any other equations containing $|A(r_n, f)|^2$ and $\Delta t_{k, n}$.
As in the last part of the previous section, we attempt to reproduce only variability properties based on the propagating fluctuations process rather than full spectral timing.
We keep those parameters fixed which are fixed in the previous fit.

Even with all these additional effects, the model is still not capable of matching the observation data.
The damping parameter $D$ is pegged to its lower bound of zero, indicating that the damping described above is ineffective in improving the fit (\citealt{Mahmoud_2019}).
This is because 
our model assumes that the intrinsic variability has a cut-off at the local generator frequency $f_{\mathrm{gen}}(r)$, as shown in Fig.~\ref{fig:propagating_fluctuations_schematic}c.
This assumption already includes some aspects of damping.
The MRI (\citealt{Balbus_1991, Balbus_1998}) is expected to produce variability up to quite fast time-scales. However, only variability slower than the local propagation time can propagate inwards as the 
faster variability is viscously damped out (\citealt{Churazov_2001, Cowperthwaite_2014, Ingram_2016, Hogg_2016, Bollimpalli_2020, Turner_2021}).
Our assumptions about the intrinsic variability are an approximation of this physical picture.
The damping parameter being pegged to zero indicates there is no need for additional damping effects.
We did not find an improvement in the fits using separate variability amplitude between the variable disc region and hot flow region, either.

Fig.~\ref{fig:psd_phase_fit_comp} (mid-right) shows the results of a joint fit to the power spectra for 2.6--4.8~keV and 35--48~keV and the phase-lag spectrum between these energy bands by allowing $f_{\mathrm{gen}}(r) \neq f_{\mathrm{prop}}(r)$ for the hot flow. For clarity, we removed the other additional effects which did not make a difference, i.e., $F_{\mathrm{var, d}}=F_{\mathrm{var, f}}$ and $D=0$.

Although we see a slightly better match to the observed peak frequency of the phase-lag spectrum than the previous fit, the model still underestimates its amplitude. 
More importantly, we still do not solve the essential issue: the model calculates similar or larger variability for higher energy bands, inconsistent with the observation that the power spectral amplitude is larger for the lower energy band. 
\cite{Mahmoud_2018b, Mahmoud_2019} introduce more complex radial dependence for the intrinsic variability, emissivity and damping to capture energy-dependent variability properties. 
However, some assumptions involved with these complications remain to be tested.
We do not explore the complex radial structure further and conclude that those additional effects implemented here are less effective than required by the high signal-to-noise ratio data obtained by {\it Insight-HXMT}.

We note that the difficulty in reproducing the observation data here lies in joint fitting to the power spectra and phase-lag spectrum.
It is possible to reproduce power spectra for these energy bands with the current model fairly well, ignoring the phase-lag spectrum.
In this case, however, the lower-energy photons would come from inner regions, because inner regions are more variable than outer regions, and predict soft lags, which is completely in disagreement with the observed hard lags.
This points to the importance of modelling cross spectra and power spectra.


\begin{figure}
	\includegraphics[width=1.\columnwidth]{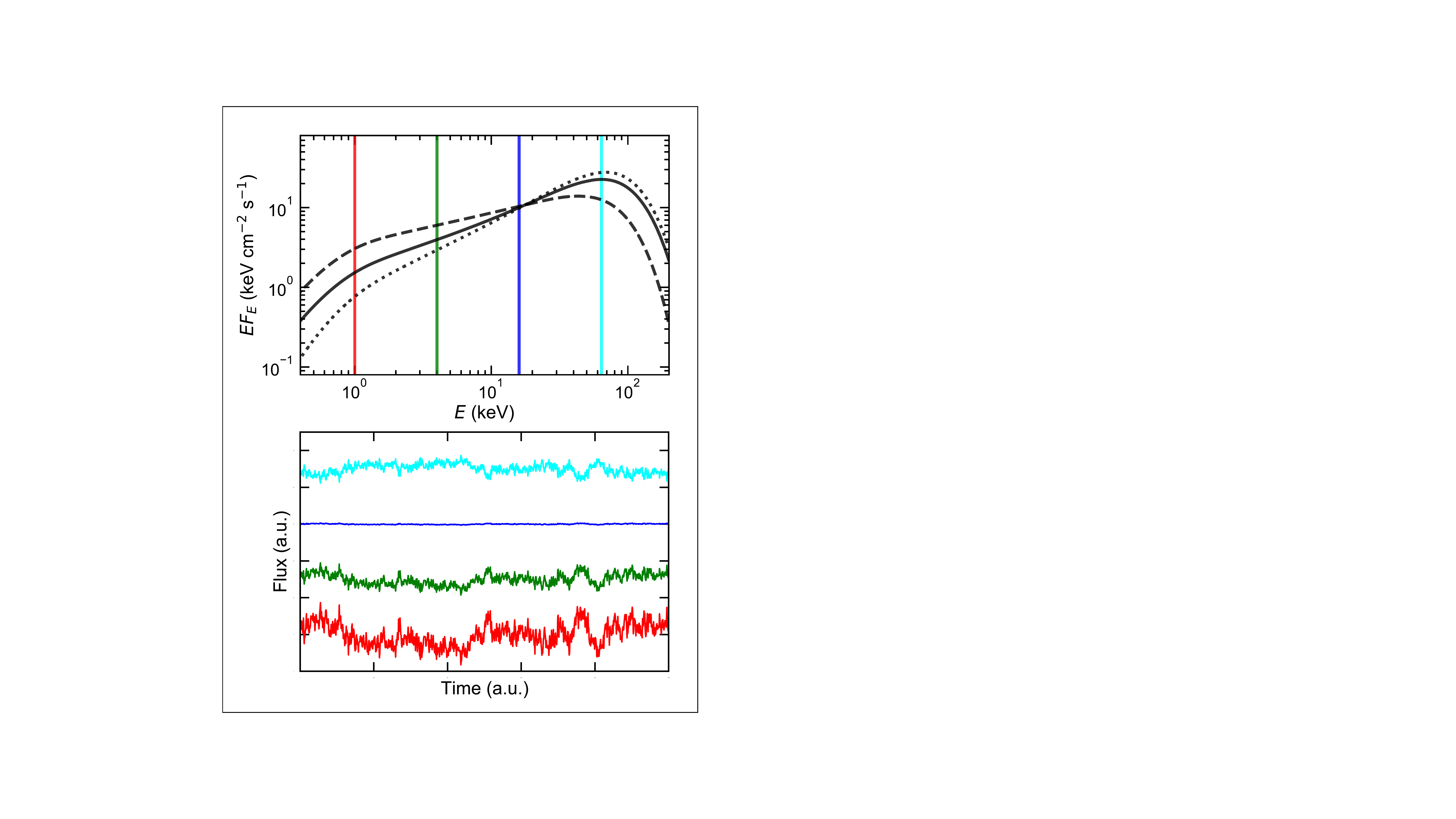}
	\caption{Schematic picture of the effect of spectral pivoting as implemented here. 
		{\it Top}: Instant local energy spectra when the mass accretion rate at the corresponding radius is higher than, equal to, and lower than the average (dashed, solid, and dotted, respectively).
		{\it Bottom}: Light curves of the local flux for different energies, $1\,\si{keV}$ (red), $4\,\si{keV}$ (green),\,$16\,\si{keV}$ (blue),\,$64\,\si{keV}$ (cyan) as marked in the top panel.
		Each light curve is normalised by its average and offset for clarity.
		The light curves in the lower energy bands (red and green) are positively correlated with the local mass accretion rate, but the fluctuations have lower amplitude as the energy increases, going to zero at the pivot point at 16~keV, and then switching to negative correlation at higher energies (cyan).
	}
	\label{fig:schematic_spectral_pivot}
\end{figure}

\begin{figure}
	\includegraphics[width=\columnwidth]{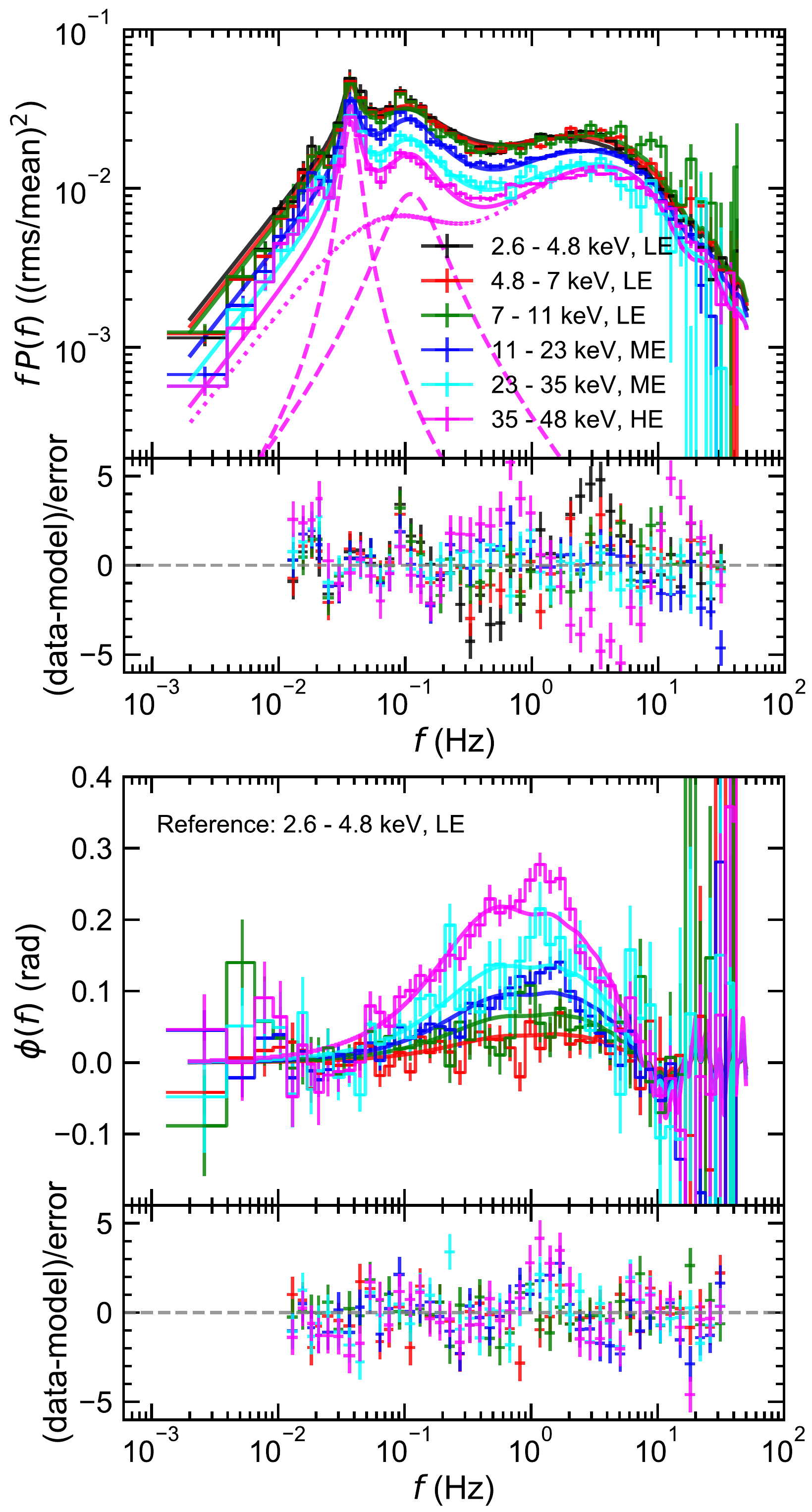}
	\caption{Joint fit to six power spectra (top) and five phase-lag spectra (bottom) across 2.6--48~keV with our new model including the spectral pivoting.
	We include two Lorentzian functions for the QPO (dashed) and harmonic (dotted). 
	In the calculation of phase-lag spectra, the lowest band of 2.6--4.8~keV is chosen as the reference band.
	The lower plot for each panel is the difference between data and model divided by one-sigma errors.
	The new model including the spectral pivot successfully reproduces all the timing data across this bandpass.
	}
	\label{fig:psd_phase_fit_multi}
\end{figure}

\subsection{Spectral pivoting}
\label{sec:rev2}

So far, we have assumed that the spectral shape of each component does not vary in time.
However, this is unphysical because mass accretion rate fluctuations make spectral parameters, e.g., the optical depth and electron temperature, vary on short time-scales (\citealt{Malzac_2003, Gandhi_2008, Yamada_2013, Bhargava_2022}).
This oversimplification limits the model's flexibility to reproduce energy-dependent variability data.
Hence we now allow the spectral shapes to fluctuate (\citealt{Veledina_2016, Veledina_2018, Mastroserio_2018, Mastroserio_2019, Mastroserio_2021}), along with their amplitude.
The schematic picture of the spectral pivoting is shown in Fig.~\ref{fig:schematic_spectral_pivot} (top).

Here, we give concise explanations of how the spectral pivoting is implemented and what the model gets to be able to handle with this update.
More detailed formalism is found in Appendix~\ref{sec:app_pivot}.
A constant spectral shape means that the spectrum at every energy reacts to mass accretion fluctuations in the same way.
We consider the mass accretion rate and energy spectrum at a certain radius. 
By defining the average and difference from the average as $\dot{m}_0$ and $\Delta \dot{m}(t)$ for the mass accretion rate and as $S_0 (E)$ and $\Delta S(E, t)$ for the spectrum, the constant spectral shape is equivalent to $\Delta S (E, t)/S_0 (E)=\Delta \dot{m} (t)/\dot{m}_0$, which is independent of energy $E$.
To let the spectral shape vary in time, we give the spectrum sensitivity to $\Delta \dot{m}(t)$ as a function of energy, $\eta (E)$, and redefine $\Delta S (E, t)/S_0 (E)=\eta(E) \Delta \dot{m} (t)/\dot{m}_0$, which now depends on energy.
The amplitude of sensitivity parameter $|\eta (E)|$ regulates how sensitive the spectrum is to a change in the mass accretion rate from its average, while its sign determines whether the spectrum reacts positively or negatively.
The spectrum gets higher (lower) with an increase in mass accretion rate if $\eta (E) >0$ ($<0$).
The energy at which $\eta (E)$ crosses zero, called the pivoting point, does not react to a change in mass accretion rate.
Light curves of local flux for different energies are illustrated in Fig.~\ref{fig:schematic_spectral_pivot} (bottom).
We note that we do not simulate light curves in the model calculations.
The decrease in $\eta (E)$ with energy, i.e., the spectrum being less sensitive to $\Delta \dot{m}(t)$ for higher energies, could let the power spectrum decrease with energy, as observed for MAXI~J1820+070, even if the mass accretion rate is more variable for central regions emitting higher-energy photons.
In our implementation, there arises no lag between different energies from the spectral pivoting itself except for the phase lag of $\pi$ when $\eta (E_1)\eta (E_2)<0$.
Our new model shares this feature of spectral pivoting with the model developed by \cite{Veledina_2016, Veledina_2018}.
The new model returns to the previous one by setting $\eta (E)=1$.

Each spectral component is expected to show its own sensitivity pattern.
We give the sensitivity parameter to each spectral component, $\eta _{\mathrm{Y}}(E)\,(\mathrm{Y}={\mathrm{d, s, h}})$, where the lower subscripts stand the variable disc, soft Comptonisation, and hard Comptonisation, respectively.
With the implementation of spectral pivoting, all $S_{\mathrm{Y}}(E)\,(\mathrm{Y}={\mathrm{d, s, h}})$ contained in the analytic expressions of power spectra and cross spectra is replaced by $\eta _{\mathrm{Y}}(E)S_{\mathrm{Y}}(E)$ (see Appendix~\ref{sec:app_pivot} for the derivation).
This means that the model's flexibility is not bound by the constraint (\ref{eq:spec_constraint}) anymore because time-averaged spectra always appear as the product with their sensitivity.
In addition, $\eta _{\mathrm{Y}}(E)S_{\mathrm{Y}}(E)$ can be negative in contrast to $0\leq S_{\mathrm{Y}}(E) \leq 1$.
The spectral pivoting gives freedom to the model in this way.

We attempt to fit the variability properties with the new model.
We have $\eta _{\mathrm{Y}}(E) S_{\mathrm{Y}}(E)$ as model parameters, instead of $S_{\mathrm{Y}}(E)$.
The negligible disc emission $S_{\mathrm{d}}(E)=0$ results in $\eta_{\mathrm{d}}(E)S_{\mathrm{d}}(E)=0$.
We fix $D=0$, in which all intrinsic variability propagates inwards without any loss.
We also fix $F_{\mathrm{var, d}}=F_{\mathrm{var, f}}$ to the typical value of $0.8$ because the sensitivity parameter $\eta (E)$ can regulate the variability amplitude.

The simultaneous fit to the power spectra for 2.6--4.8~keV and 35--48~keV and the phase-lag spectrum between these energy bands with the new model is shown in Fig.~\ref{fig:psd_phase_fit_comp} (right).
We see significant improvement in variability modelling by allowing the spectral shapes to vary in time. 
Our new model captures the energy-dependent variability, pointing to the importance of spectral pivoting in modelling variability at high energies.

To study the variability for a continuous energy range, we split energy between 2.6--4.8~keV (LE) and 35--48~keV (HE) into four bands, i.e., 4.8--7~keV (LE), 7--11~keV (LE), 11--23~keV (ME), 23--35~keV (ME), where the telescopes used are specified in parenthesis, and attempt to reproduce power spectra for these six energy bands and phase-lag spectra with respect to the lowest energy band for the rest of five energy bands. 
We minimize 
\begin{equation}
	\begin{split}
		&\sum_{j, k} \left( \frac{P_{\mathrm{data}}(E_{j}, f_{k})-P_{\mathrm{model}}(E_{j}, f_{k})}{\Delta P_{\mathrm{data}}(E_{j}, f_{k})} \right) ^2 \\
		+ &\sum _{\substack{j, k\\(E_{j} \neq E_{\mathrm{r}})}} \left( \frac{\phi_{\mathrm{data}}(E_{\mathrm{r}}, E_{j}, f_{k})-\phi_{\mathrm{model}}(E_{\mathrm{r}}, E_{j}, f_{k})}{\Delta \phi_{\mathrm{data}}(E_{\mathrm{r}}, E_{\mathrm{j}}, f_{k})} \right) ^2 , \\ 
	\end{split}
\end{equation}
through the fit, where $E_j$ is each energy band and $E_{\mathrm{r}}=2.6\textrm{--}4.8\,\si{keV}$ the reference band.
For more complete modelling, we add two Lorentzian functions to model the QPOs in power spectra by using the {\tt XSPEC} model {\tt lorentz}.
To fix the centroid frequency and width of the QPO models, we perform a phenomenological fit to power spectra with the sum of four Lorentzian functions, where two of them are used to model each bump of the broad-band variability.
We then extract the centroid and width of $(3.66 \times 10^{-2}\,\si{Hz},\,1.20 \times 10^{-2}\,\si{Hz})$ for the QPO fundamental, and $(9.44 \times 10^{-2}\,\si{Hz},\,1.16 \times 10^{-1}\,\si{Hz})$ for the second harmonic as typical values. 
Thus, the ${\tt lorentz}$ model has only one free parameter, the normalisation.
On the other hand, we do not use any additional models in phase-lag spectra due to the relatively small QPO features.
The results of the joint fit to six power spectra and five phase-lag spectra are shown in Fig.~\ref{fig:psd_phase_fit_multi}.
Each component forming power spectra is explicitly plotted with dashed (QPOs) and dotted (broad-band) lines only for the highest energy band of 35--48~keV.
Model parameter values are summarized in Table~\ref{tab:model_parameter_values_multi}. 

We find that the new model matches observations well for all energy bands whilst keeping parameter values similar to those found in the joint fitting for 2.4--4.8~keV and 35--48~keV only (Fig.~\ref{fig:psd_phase_fit_comp} (right)).
It is interesting to note that the spectral parameter for the soft Comptonisation component $\mu _{\mathrm{s}}(E) S_{\mathrm{s}}(E)$ decreases with energy and finally reaches a negative value at the highest energy band of 35--48~keV.
This means that the soft Comptonisation component increases for an increase in mass accretion rate at low energies ($\lesssim 35\,\si{keV}$), whereas it decreases at high energies ($\gtrsim 35\,\si{keV}$), showing the pivoting point of $\sim 35\,\si{keV}$.

Although the broad-band variability has been studied with {\it Insight-HXMT} observations (e.g. \citealt{Wang_2020, Yang_2022}), we succeeded in reproducing it with a physically-motivated model for the first time.
In addition, while propagating fluctuations models have been applied up to $\sim 35\,\si{keV}$ (\citealt{Mahmoud_2018, Mahmoud_2018b}) with {\it RXTE} observations, we extend the energy range up to 48~keV using {\it Insight-HXMT} observations with significantly improved residuals.
Our successful modelling shows the propagating fluctuations scenario holds good up to high energy bands, keeping it the most plausible explanation for the aperiodic variability.

\begin{table}
\caption{Model parameter values used in Figs.~\ref{fig:psd_phase_fit_multi}. 
Fixed parameter values are the same as in Table~\ref{tab:model_parameter_values}.}
\begin{tabular}{ll}
\hline
Symbol&Value\\
\hline
$F_{\mathrm{var, d}}$                               &$0.8$ (f)                        \\
$B_{\mathrm{f}}$                                    &$401$                            \\
$m_{\mathrm{f}}$                                    &$2.59$                           \\
$B^{(\mathrm{p})}_{\mathrm{f}}$                     &$108$                            \\
$\eta_{\mathrm{s}} S_\mathrm{s}(2.6\textrm{--}4.8\,\si{keV})$         &$0.315$                          \\
$\eta_{\mathrm{h}} S_\mathrm{h}(2.6\textrm{--}4.8\,\si{keV})$         &$0.588$                        \\
$\eta_{\mathrm{s}} S_\mathrm{s}(4.8\textrm{--}7\,\si{keV})$           &$0.242$                        \\
$\eta_{\mathrm{h}} S_\mathrm{h}(4.8\textrm{--}7\,\si{keV})$           &$0.601$                        \\
$\eta_{\mathrm{s}} S_\mathrm{s}(7\textrm{--}11\,\si{keV})$            &$0.189$                        \\
$\eta_{\mathrm{h}} S_\mathrm{h}(7\textrm{--}11\,\si{keV})$            &$0.617$                        \\
$\eta_{\mathrm{s}} S_\mathrm{s}(11\textrm{--}23\,\si{keV})$           &$0.116$                        \\
$\eta_{\mathrm{h}} S_\mathrm{h}(11\textrm{--}23\,\si{keV})$           &$0.549$                        \\
$\eta_{\mathrm{s}} S_\mathrm{s}(23\textrm{--}35\,\si{keV})$           &$0.049$                        \\
$\eta_{\mathrm{h}} S_\mathrm{h}(23\textrm{--}35\,\si{keV})$           &$0.498$                        \\
$\eta_{\mathrm{s}} S_\mathrm{s}(35\textrm{--}48\,\si{keV})$           &$-0.052$                       \\
$\eta_{\mathrm{h}} S_\mathrm{h}(35\textrm{--}48\,\si{keV})$           &$0.484$                        \\
\hline
$\chi ^2 /\mathrm{d.o.f.}$                          &$1095.0/457$ \\
\hline
\end{tabular}\\
\label{tab:model_parameter_values_multi}
\end{table}


\section{Joint spectral-timing fit with spectral pivoting}
\label{sec:spec_timing}

\begin{figure*}
	\includegraphics[width=\linewidth]{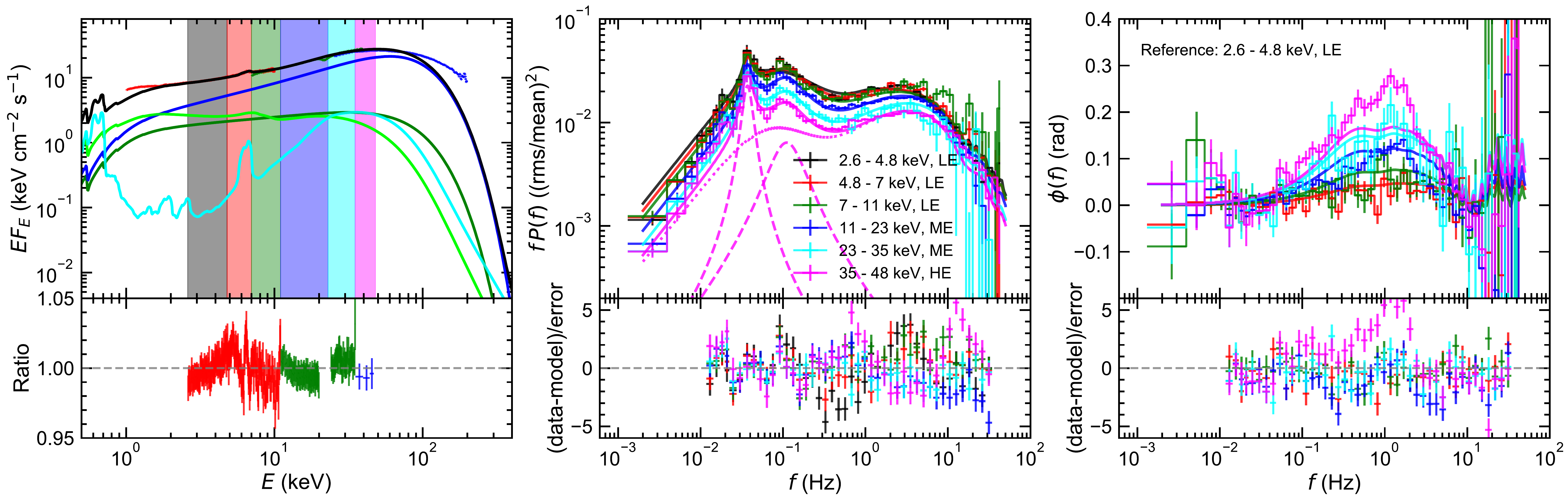}
	\caption{Spectral-timing fit to the time-averaged energy spectrum (left), six power spectra (middle), and five phase-lag spectra (right) across 2.6--48~keV with our new model including spectral pivoting.
	In the left panel, the soft Comptonisation and its associated reflection are plotted with the green and light green lines, while the hard Comptonisation and its associated reflection are plotted with the blue and light blue lines.
	The black line shows their sum. We include the effect of 
	galactic absorption of $N_{\mathrm{H}}=1.4 \times 10^{21}\,\si{cm^{-2}}$ on the spectral components. 
	The colours of the shaded regions in the energy spectral plot show the energy band used in the power spectra and phase-lag spectra.
	In the mid panel, the power spectrum at the highest energy band includes the QPO and harmonic, as in Fig.~\ref{fig:psd_phase_fit_multi}.
	The bottom panels show residuals. 
	The data-to-model ratio is used for the energy spectrum, while the difference between data and model divided by one-sigma errors is used for power spectra and phase-lag spectra.
	We successfully fit all the data in this bandpass with our updated model including the spectral pivoting, except for a slight underestimate of the phase-lag spectrum around the peak for the highest energy band.
	}
	\label{fig:spec_psd_phase_fit}
\end{figure*}

\begin{figure}
	\includegraphics[width=\columnwidth]{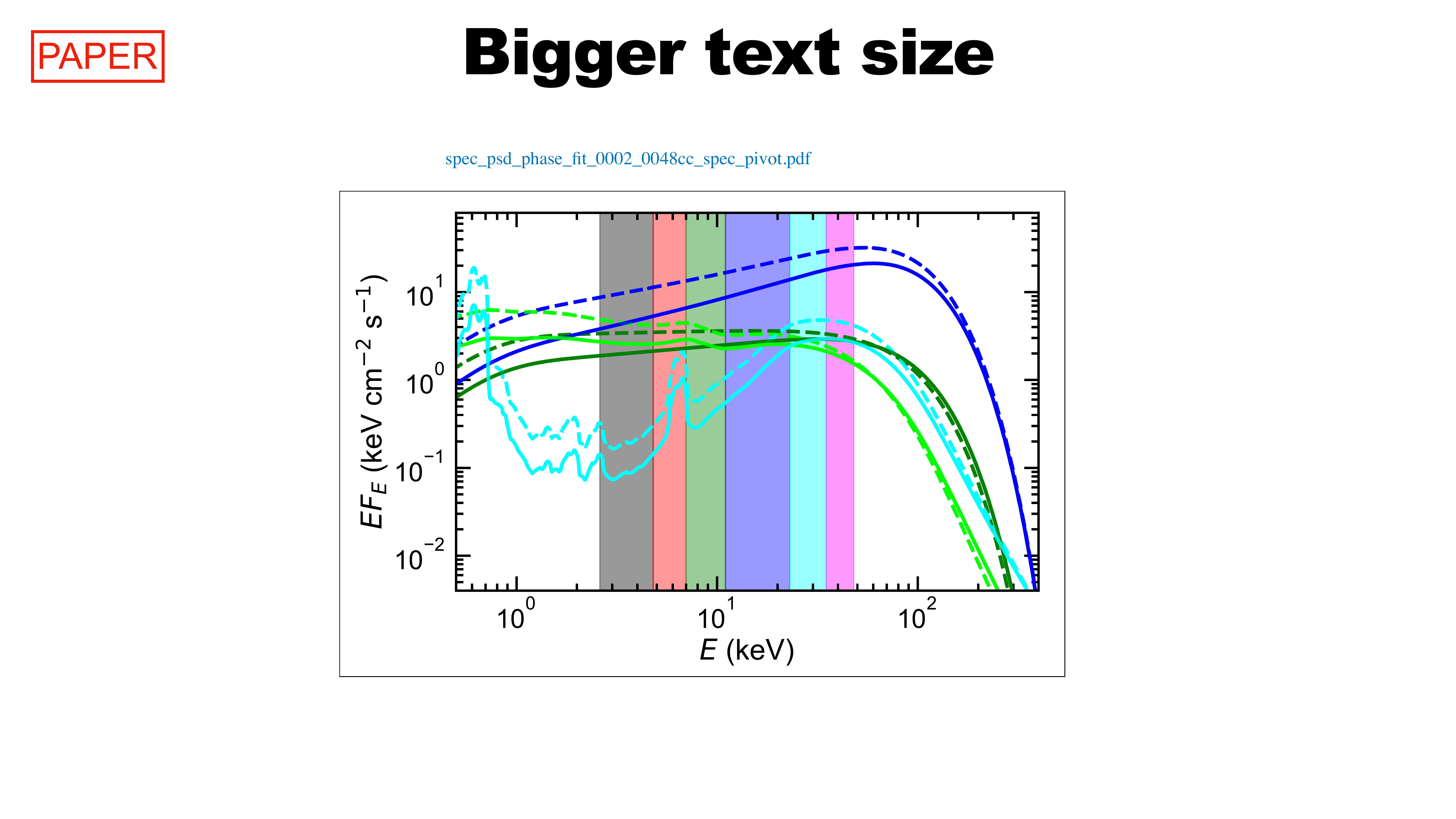}
	\caption{Changes in the shape of the energy spectral components due to changes in mass accretion rate derived from the spectral-timing fit in Fig.~\ref{fig:spec_psd_phase_fit}. The average energy spectrum for the soft (green) and hard (blue) Comptonisation components and their reflected emission (pale green and cyan, respectively) are shown as solid lines, while the dashed lines show the effect of doubling the mass accretion rate. 
	}
	\label{fig:spectral_pivoting_fit_reflection}
\end{figure}

\begin{table}
\caption{Model parameter values derived from the joint spectral-timing fit in Fig.~\ref{fig:spec_psd_phase_fit}.
Fixed parameters related to the spectrum are 
the seed photon temperature $kT_{\mathrm{seed, s}}=kT_{\mathrm{seed, h}}=0.2\,\si{keV}$,
electron temperature $kT_{\mathrm{e, s}}=kT_{\mathrm{e, h}}=23\,\si{keV}$,
Fe abundance $Z_{\mathrm{Fe}}=1.1$, 
inclination angle $i=66^{\circ}$, 
black hole spin $a^{*}=0$,
electron density $N_{\mathrm{e}}=10 ^{20}\,\si{cm^{-3}}$, 
inner radius of reflection region $R_{\mathrm{in, s}}=45R_{\mathrm{g}}$, and
outer radii of reflection region $R_{\mathrm{out, s}}=R_{\mathrm{out, h}}=1000R_{\mathrm{g}}$.
The lower subscript `$\mathrm{s}$' (`$\mathrm{h}$') denotes the soft (hard) Comptonisation or its associated reflection component.
Fixed parameters related to variability are the same as in Table~\ref{tab:model_parameter_values} in addition to the extra parameters about reverberation, $t_{0, \mathrm{s}}=t_{0, \mathrm{h}}=6\,\si{ms}$ and $\Delta t_{0, \mathrm{s}}=\Delta t_{0, \mathrm{h}}=10\,\si{ms}$.
}
\begin{tabular}{llll}
\hline
Component&Model&Symbol&Value\\
\hline
\multicolumn{4}{c}{Spectral parameters}\\
\hline
Soft Comptonisation&{\tt nthcomp}                               &$\Gamma_{\mathrm{s}}$          &$1.81$\\
and reflection     &                                            &$\mathrm{norm}_{\mathrm{s}}$   &$1.38$\\
                   &{\tt relxillCp}                             &$\log_{10} \xi _{\mathrm{s}}$  &$3.44$\\
                   &                                            &$\mathrm{norm}^{(\mathrm{r})}_{\mathrm{s}}$&$0.0395$\\
Hard Comptonisation&{\tt nthcomp}                               &$\Gamma_{\mathrm{h}}$          &$1.50$\\
and reflection     &                                            &$\mathrm{norm}_{\mathrm{h}}$   &$2.11$\\
                   &{\tt relxillCp}                             &$R_{\mathrm{in, h}}$  &$78$\\
                   &                                            &$\log_{10} \xi _{\mathrm{h}}$  &$1.70$\\
                   &                                            &$\mathrm{norm}^{(\mathrm{r})}_{\mathrm{h}}$&$0.0328$\\
\hline
\multicolumn{4}{c}{Variability parameters}\\
\hline
Broad-band         &{\tt our model}                             &$B_{\mathrm{f}}$                &$862$\\
                   &                                            &$m_{\mathrm{f}}$                &$2.81$\\
                   &                                            &$B^{(\mathrm{p})}_{\mathrm{f}}$ &$189$\\
                   &                                            &$\eta_{\mathrm{0, s}}$          &$1.023$\\
                   &                                            &$\eta_{\mathrm{1, s}}$          &$-0.568$\\
                   &                                            &$\eta_{\mathrm{0, h}}$          &$1.527$\\
                   &                                            &$\eta_{\mathrm{1, h}}$          &$-0.580$\\
\hline
                   &                                            &$\chi ^2/\mathrm{d.o.f.}$       &$1993.1/1795$\\
\hline
\end{tabular}\\
\label{tab:spec_timing_parameter_values}
\end{table}

We come back to implement a full spectral-timing analysis, rather than just the series of timing analyses above.
We attempt to fit the energy-dependent timing properties from 2.6--48~keV along with the time-average energy spectrum at the corresponding energy range by minimizing
\begin{equation}
	\begin{split}
		&\sum_{i} \left( \frac{S_{\mathrm{data}}(E_{i})-S_{\mathrm{model}}(E_{i})}{\Delta S_{\mathrm{data}}(E_{i})} \right) ^2 \\
		+ &\sum_{j, k} \left( \frac{P_{\mathrm{data}}(E_{j}, f_{k})-P_{\mathrm{model}}(E_{j}, f_{k})}{\Delta P_{\mathrm{data}}(E_{j}, f_{k})} \right) ^2 \\
		+ &\sum _{\substack{j, k\\(E_{j} \neq E_{\mathrm{r}})}} \left( \frac{\phi_{\mathrm{data}}(E_{\mathrm{r}}, E_{j}, f_{k})-\phi_{\mathrm{model}}(E_{\mathrm{r}}, E_{j}, f_{k})}{\Delta \phi_{\mathrm{data}}(E_{\mathrm{r}}, E_{\mathrm{j}}, f_{k})} \right) ^2 , \\ 
	\end{split}
\end{equation}
where $S_{\mathrm{data}}(E)$, $\Delta S_{\mathrm{data}}(E)$ are the observed time-averaged spectrum and its one-sigma error, $S_{\mathrm{model}}(E)$ the modelled time-averaged spectrum, and $E_{i}$ each energy bin in the time-averaged spectrum.
We remove clear calibration features seen in the ME spectrum for 20--24 keV (light green regions in Fig.~\ref{fig:observation_data} (top)) from the spectral modelling.

To model the energy spectrum, we account for not only the soft and hard Comptonisation components but their disc reflection.
We ignore emission from the turbulent disc due to its negligible contribution above the lowest energy of 2.6~keV (few \% at 2.6~keV in the spectral fit found in K22).
We also ignore the negligible effect of galactic absorption.
We use the {\tt XSPEC} model {\tt nthcomp} (\citealt{Zdziarski_1996, Zycki_1999}) for the Comptonisation components, and {\tt relxillCp} provided in {\tt relxill} version 2.0 (\citealt{Garcia_2014, Dauser_2022}) for the reflected components.
Finally, we use 
\begin{equation}
    ({\tt nthcomp}+{\tt relxillCp}) + ({\tt nthcomp}+{\tt relxillCp}),
\end{equation} 
where each bracket corresponds to the soft Comptonisation/reflection and hard Comptonisation/reflection, respectively.

To connect the time-averaged spectrum and variability consistently, we take reverberation into account in our timing model.
Its implementation is updated from that in K22 mainly due to the inclusion of spectral pivoting.
We summarize how the reverberation behaves in our new model here, while the detailed formalism is described in Appendix~\ref{sec:app_reflection}.

The illuminating Comptonisation spectrum changing its shape with time results in the reflected spectrum also changing its shape with time.
As in the previous section, we consider a certain radius.
Along with the mass accretion rate and direct emission, we account for the reflected emission associated with the direct emission.
Defining the average and difference from it as $S^{(\mathrm{r})}_{0}(E)$ and $\Delta S^{(\mathrm{r})}(E, t)$, we assume $\Delta S^{(\mathrm{r})}(E, t)/S^{(\mathrm{r})}_{0}(E) = (\Delta S(E, t)/S_{0}(E))\otimes h(t)=\eta (E) (\Delta \dot{m}_{0}(t)/\dot{m}_{0})\otimes h(t)$.
We use the upper script `${(\mathrm{r})}$' to stand for the reflected emission.
The convolution in time is denoted by $\otimes$, and $h(t)$ is called the impulse response, which is the time evolution of reflected emission for a flash of illumination.
All information as to the disc response, such as the delay for the direct emission due to an additional light crossing path and the duration due to the different delay times for different locations of reflection, are encoded in $h(t)$. 

The relation of spectral variation between the direct and reflected emission means that the reflected emission follows variations of the direct emission at the corresponding energy with some time delay, as long as the variability is slow enough not to be washed out via reprocessing, i.e., via the operation of the convolution.
In the simple case of $h(t)=\delta (t-\tau)$, variations of the reflected emission exactly lag behind those of the direct emission with the time delay of $\tau$: $\Delta S^{(\mathrm{r})}(E, t)/S^{(\mathrm{r})}_{0}(E)=\Delta S(E, t-\tau )/S_{0}(E)$.

Each reflected component has its own impulse response as each Comptonisation component illuminates different parts of the accretion flow (\citealt{Zdziarski_2021}).
We define the impulse response with a top-hat function:
\begin{equation}
    h_{\mathrm{Y}} (t)=
    \begin{cases}
        1/\Delta t_{0, \mathrm{Y}} & (|t-t_{0, \mathrm{Y}}|< \Delta t_{0, \mathrm{Y}}/2)  ,\\
        0 & (\mathrm{othewise}),
    \end{cases}
\end{equation}
where $\mathrm{Y}={\mathrm{s, h}}$ are associated with the soft and hard Comptonisation components, respectively.
The parameters $t_{0, \mathrm{Y}}$ and $\Delta t_{0, \mathrm{Y}}$ characterize the delay and duration, respectively. 
More realistic impulse responses are required, especially for low energy bands $E\lesssim 2\,\si{keV}$, where the quasi-thermal emission due to the reprocessing dominates high-frequency variability ($\gtrsim 1\,\si{Hz}$).
However, the top-hat function appears to be a good approximation for high energies, where Comptonisation largely determines variability properties.

For the consistency between the spectral modelling and variability modelling, we calculate the fractional time-averaged spectra required in our timing model, $S_{\mathrm{d}}(E)(=0)$, $S_{\mathrm{s}}(E)$, $S_{\mathrm{h}}(E)$, $S^{(\mathrm{r})}_{\mathrm{s}}(E)$, $S^{(\mathrm{r})}_{\mathrm{h}}(E)$, from the spectral models, {\tt nthcomp} and {\tt relxillCp}.
To obtain $\eta _{\mathrm{s}}(E)S_{\mathrm{s}}(E)$, $\eta _{\mathrm{h}}(E)S_{\mathrm{h}}(E)$, $\eta _{\mathrm{s}}(E)S ^{(\mathrm{r})} _{\mathrm{s}}(E)$, and $\eta _{\mathrm{h}}(E)S ^{(\mathrm{r})} _{\mathrm{s}}(E)$, it is simple to assume a function for the sensitivity parameters $\eta _{\mathrm{Y}}(E)\,(\mathrm{Y}={\mathrm{s, h}})$.
We note that we do not need $\eta_{\mathrm{d}}(E)$ due to $S_{\mathrm{d}}(E)=0$.
Given that $\eta _{\mathrm{Y}}(E)$ can switch its sign at the pivoting point and that the fractional rms of the broad-band variability roughly changes with energy logarithmically (\citealt{Gierlinski_2005, Yang_2022}), it is fair to make a phenomenological assumption of 
\begin{equation}
    \eta _{\mathrm{Y}}(E)=\eta_{0, \mathrm{Y}} + \eta_{1, \mathrm{Y}} \log _{10} \left( E\,[\si{keV}] \right). 
    \label{eq:eta_phenomenology}
\end{equation}
The model parameter $\eta_{0, \mathrm{Y}}$ is the sensitivity at $1\,\si{keV}$, while $\eta _{1, \mathrm{Y}}$ determines its gradient to energy.

We note the difference in the model calculations between the timing fits (Section~\ref{sec:rev2}) and spectral-timing fits.
In the timing fits, $\eta(E) S_{0}(E)$ is a model parameter, and it is impossible to disentangle this product.
On the other hand, $S_{0}(E)$ and $\eta (E)$ are separately modelled in the spectral-timing fits.
The former is calculated from spectral models, the latter is from equation (\ref{eq:eta_phenomenology}).

In the joint spectral-timing fit, we fix the seed photon temperature of Comptonisation components to the typical disc temperature in this state, $kT_{\mathrm{seed, s}}=kT_{\mathrm{seed, h}}=0.2\,\si{keV}$ (\citealt{DeMarco_2021}, K22).
Since the electron temperature is difficult to constrain from the energy band of interest, we fix it to $kT_{\mathrm{e, s}}=kT_{\mathrm{e, h}}=23\,\si{keV}$, as in K22.
While we allow the inner radius of the reflector for the hard Comptonisation component to be free, we fix that for the soft Comptonisation component to $R_{\mathrm{in, s}}=45R_{\mathrm{g}}$ corresponding to the outer edge of the variable flow located at $r_{\mathrm{out}}=45$.
Following K22, we fix the inclination angle to $i=66^{\circ}$ (\citealt{Torres_2020}) and Fe abundance to $Z_{\mathrm{Fe}}=1.1$.
We also set the black hole spin to $a^{*}=0$, consistent with $r_{\mathrm{in}}=6$ in the timing model, and use the high electron density of $N_{\mathrm{e}}=10^{20}\,\si{cm^{-3}}$ (\citealt{Garcia_2016, Mastroserio_2021}).
The delay and duration of the impulse response are, in principle, derived from the location and geometry of illuminating source and reflector.
However, in the geometry assumed, the time-scales of reverberation $\lesssim 10\,\si{ms}$ (corresponding to the light crossing of $\lesssim 250R_{\mathrm{g}}$ for $M_{\mathrm{BH}}=8M_{\odot}$) are shorter than variability time-scales of interest ($20\,\si{ms}\textrm{--}100\,\si{s}$).
In addition, reverberation signatures are unclear across the energy bands of interest ($2.6\textrm{--}48\,\si{keV}$), and small alterations of the impulse response due to small changes of the accretion flow geometry do not significantly affect the variability properties.
Thus, we simply fix $t_{0, \mathrm{s}}=t_{0, \mathrm{h}}=6\,\si{ms}$ and $\Delta t_{0, \mathrm{s}}=\Delta t_{0, \mathrm{h}}=10\,\si{ms}$ as typical values.
The top-hat impulse response with these values appears to be good approximations of more realistic ones (K22). 

The results of simultaneous modelling of the energy spectrum, six power spectra, and five phase-lag spectra are shown in Fig.~\ref{fig:spec_psd_phase_fit}.
The comparison between the data and model is also plotted as the ratio for the energy spectrum and the difference divided by one-sigma errors for the variability.
Model parameter values are found in Table~\ref{tab:spec_timing_parameter_values}.
Overall, our new model successfully reproduces both time-averaged and variability properties, although the discrepancies are seen in the phase-lag spectrum between 35--48~keV and 2.6--4.8~keV (magenta), which is discussed in Section~\ref{sec:discussion_limitations}.
This modelling is the first simultaneous fit to spectrum and variability using our model.
The uncertainties of the derived parameter values are evaluated with a Markov Chain Monte Carlo (MCMC) analysis in Appendix~\ref{sec:app_mcmc}.

The spectral variation derived from the fit is shown in Fig.~\ref{fig:spectral_pivoting_fit_reflection}.
The spectra for the mass accretion rate being its average and double the average are plotted with solid and dashed lines.
For illustration purposes, we ignore all effects from the impulse response for reverberation, such as time delay, i.e., we assume $h(t)=\delta (t)$.
This means that the Comptonisation and its associated reflection behave completely in the same way, $\Delta S(E, t)/S_{0}(E)=\Delta S^{(\mathrm{r})}(E, t)/S^{(\mathrm{r})}_{0}(E)$.
Generally, all spectra are less sensitive for higher energies to mass accretion rate fluctuations, which results in a decrease in the power spectrum with energy.
We see the pivoting point at $\sim 50\,\si{keV}$ for the soft Comptonisation and its reflection, which roughly agrees with that at $\sim 35\,\si{keV}$ derived from the fit only to the timing properties in the previous section.


\section{Discussion}
\label{sec:discussion}


\begin{figure}
	\includegraphics[width=\linewidth]{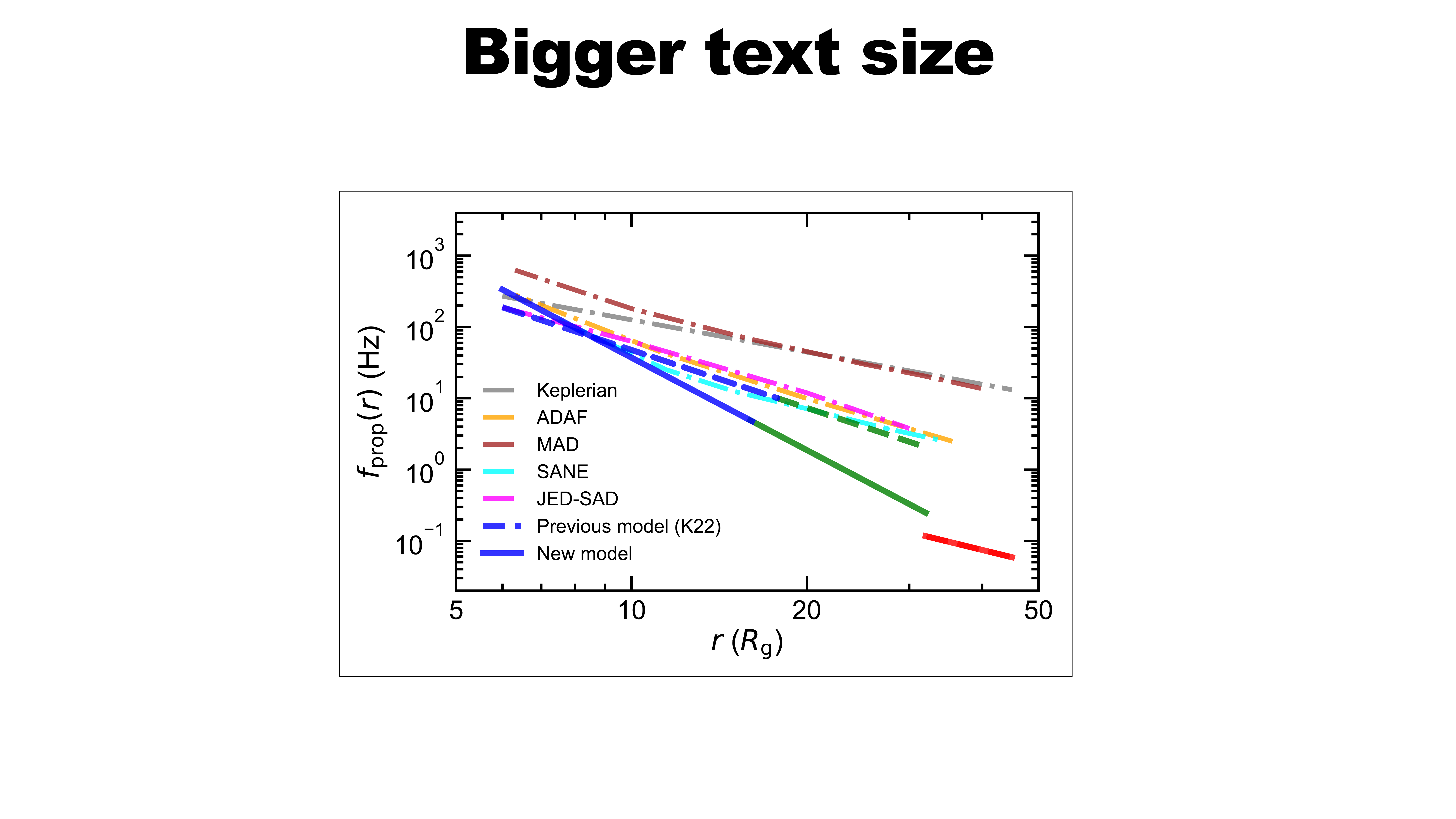}
	\caption{Propagation frequency (solid) as a function of radius derived from the spectral-timing fit.
	Red, green, and blue colours denote the variable disc, soft Comptonisation, and hard Comptonisation regions, respectively.
	The propagation frequency derived from our previous work and predicted from theoretical models are plotted with a dashed line and dash-dotted lines (see K22 for details).
	A black hole mass of $M_{\mathrm{BH}}=8M_{\odot}$ is assumed.
	The Keplerian frequency is also plotted with a grey dash-dotted line for reference. We caution that none of these models except the JED-SAD has an explicit transition from the flow to the disc.
	}
	\label{fig:fvisc_updates}
\end{figure}

\begin{figure}
	\includegraphics[width=\columnwidth]{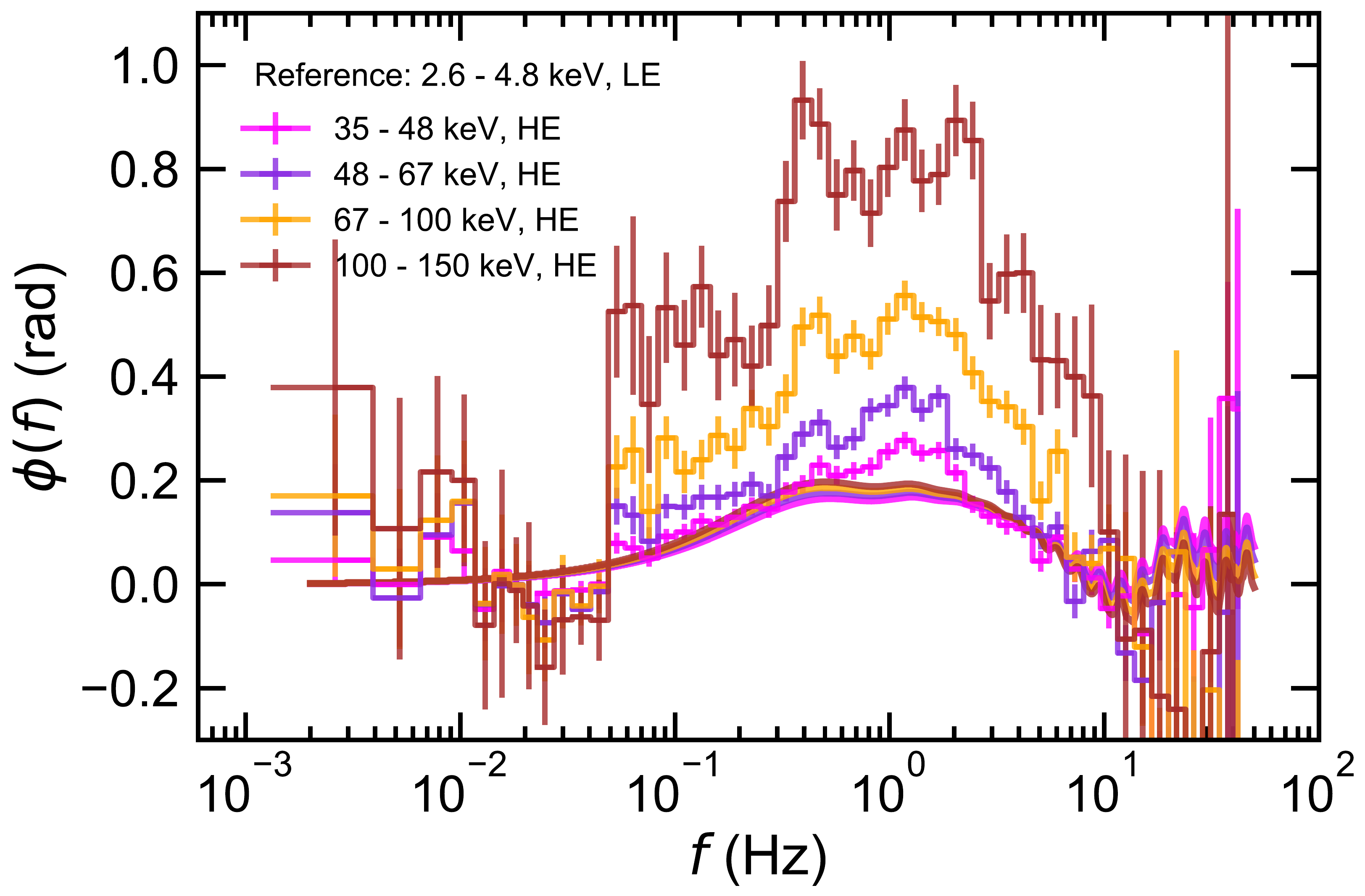}
	\caption{The predicted high energy phase lags (lines) versus the data (points), calculated using the standard 2.6-5.8~keV reference band. The lowest energy band on this plot is 35--48~keV (magenta), which is the highest energy band included in the spectral-timing fits. The predicted phase lags at higher energies saturate beyond 40~keV, unlike the data, which show a clear increase in phase lag with energy, pointing out the limitations of the model. 
	}
	\label{fig:phase_extrapolation}
\end{figure}


\subsection{Generator time-scale and propagation time-scale}

The characteristic time-scales on which the fluctuations are propagated at each radius are derived from the spectral-timing fit (Fig.~\ref{fig:spec_psd_phase_fit}).
We compare the propagation frequency derived (solid) to those predicted by different hot flow models (dash-dotted) in Fig.~\ref{fig:fvisc_updates}. 
The theoretical propagation frequencies for the Advection Dominated Accretion Flow (ADAF; \citealt{Narayan_1997}), Standard And Normal Evolution accretion flow (SANE; \citealt{Narayan_2012}), Magnetically Arrested Disc (MAD; \citealt{Narayan_2012}), and Jet Emitting Disc (JAD; \citealt{Marcel_2018}) are calculated in a standard way by dividing the accretion speed by radius assuming $M_{\mathrm{BH}}=8M_{\odot}$ (see K22 for details). 
The propagation frequency derived from our previous model from K22 is also plotted (dashed).

In K22, we found fairly good agreement of the derived propagation time-scales with those in ADAF, SANE, and JED rather than MAD. 
The propagation time-scale derived from our new model is now not very similar to any theoretical predictions.
Here, allowing the generator and propagation frequencies to be different makes their radial dependence steeper, i.e., from $f_{\mathrm{prop}}(r)\propto r^{-2.7}$ to $f_{\mathrm{prop}}(r)\propto r^{-4.22}$.
This steep radial dependence is required to reproduce both the observed large phase lags and high-frequency broad-band variability simultaneously.
Indeed, the propagation time-scale derived is robust against uncertainties of the relationship between the generator time-scale and propagation time-scale.
Using our new model including the spectral pivoting, the assumption of $f_{\mathrm{gen}}(r)=f_{\mathrm{prop}}(r)$ also gives a similar propagation time-scale, although the fit is not as good as that obtained in the previous section with $f_{\mathrm{gen}}(r)\neq f_{\mathrm{prop}}(r)$.
The key feature that our new model requires is a slow propagation speed enough to reproduce observed phase lags.
The propagation time-scale of MAD (brown dash-dotted line) is too short to explain the observed phase lags.
Thus, our results still prefer SANE rather than MAD, although MAXI~J1820+070 displays a powerful jet (\citealt{Bright_2020}).


\subsection{Origin of QPOs}

Our full spectral-timing modelling accounts for all the X-ray spectrum and rapid variability except for the QPOs.
We model these QPOs by adding peaked Lorentzian components in the power spectra.
No extra component is added to the phase-lag spectra simply because the QPO features are not very clear across 2.6--48~keV.
We did not add any other spectral components for the QPOs, implicitly assuming that the QPO is a modulation of the spectral components already included in the model (multiplicative) rather than being associated with an additional spectral component (additive).

Our successful modelling does not give much room for an additional emission component only related to the QPOs (Fig.~\ref{fig:spec_psd_phase_fit}), supporting the assumption above. 
This result is consistent with a QPO produced predominantly from a global mode of the flow rather than an intrinsic change in intensity with QPO frequency. 
We specifically have in mind Lense-Thirring (vertical) precession of the entire hot flow, where the observed luminosity of the Comptonisation component(s), including all their stochastic variability, are modulated by the changing projected area of the translucent hot flow as the viewing angle changes with QPO phase  (\citealt{Fragile_2007, Ingram_2009, Ingram_2011, Ingram_2012}). 
This picture agrees with the new polarisation results for Cyg~X-1 in the low/hard state, which requires the hot X-ray emitting plasma to be radially extended (\citealt{Krawczynski_2022}).
Conversely, the alternative model of a precessing jet suggested by \cite{Ma_2021} is challenged by the polarization results because it requires the hot X-ray emitting region to align with the jet.


\subsection{Limitations of our new model}
\label{sec:discussion_limitations}

Our new model gives a poor fit to both the energy spectrum and phase lags beyond $\sim 40$--$50$~keV. 
From Fig.~\ref{fig:spec_psd_phase_fit} (left), the spectral model clearly underestimates the data above $\sim$ 100~keV.
The hard Comptonisation spectrum rolls over too fast to match the observed data.
The comparison of the phase-lag spectrum between the model and data at high energies is shown in Fig.~\ref{fig:phase_extrapolation}.
The model phase lags increase smoothly with energy to $\sim 40\,\si{keV}$ (Fig.~\ref{fig:spec_psd_phase_fit} (right)) but then saturates to a constant value rather than continually increasing as in the data.
The lag behaviour arises as the fraction of the total spectrum which is made of the hard (and long lagged) Comptonisation spectrum increases up to around $\sim 40\,\si{keV}$, but after this point, the hard Comptonisation dominates, leading to the saturation of lag. 

The spectral mismatch could be fixed if there is additional stratification of the energy spectrum of the hot flow, so the very innermost regions produce an even hotter/harder Comptonisation component. 
In many ways, this is quite natural.
The two Comptonisation components used here for the spectral decomposition are only an approximation to a continuous flow with (presumably) continuous stratification, even if we do expect there physically to be two main regions. 
Close to the disc, seed photons for Comptonisation are predominantly from the disc. 
However, it is quite easy for this Comptonisation to become optically thick along the equatorial direction, shielding the inner regions from the disc photons so that seed photons are predominantly from cyclo-synchrotron (\citealt{Poutanen_2014}).
Nonetheless, there could still be some radial temperature/spectral hardness gradients in this second region which could produce additional emissions at the hardest energies (\citealt{Poutanen_2014}). 
We note that the JED models (e.g. \citealt{Marcel_2018}) also predict a continuously increasing temperature/harder spectrum with radius in their hot flow. 

However, including the additional harder Comptonisation component probably does not fully solve issues with the phase lag, as the amount of increased lag should be rather small as the propagation speed is already high. 
Yet the data show a large increase in lag between high energy bands. 
It seems more likely that there are other factors at work affecting the lags, potentially related to changing temperature in the flow.

There is another feature which is lacking from our new model.
It is a physical description of the spectral pivoting from the Comptonisation process.
Currently, the model assumes that the spectra pivot in a synchronous way, i.e., the local spectrum at every energy responds to fluctuations of the local mass accretion rate simultaneously.
Although the magnitude of the response can be different between different energies, as seen in Fig.~\ref{fig:spectral_pivoting_fit_reflection}, there is no causal connection between them.
There can be 0 or $\pi$ of phase lags arising from the spectral pivoting itself (the phase lag of $\pi$ arises if one energy band is above the pivoting point and the other band is below it, i.e., $\eta(E_1)\eta(E_2)<0$).
Thus, in our new model, the lags between different energies are still due to the propagating fluctuation process, as is the case for the model developed by \cite{Veledina_2016, Veledina_2018}.
The spectral pivoting implemented in our new model can strongly affect power spectra but has only a relatively mild effect on phase-lag spectra.
Indeed, our new model is able to reproduce energy-dependent power spectra fairly well up to $\sim 100\,\si{keV}$, even though it fails to match the phase-lag spectra.

However, the physical picture of Comptonisation described above should give a characteristic spectral pivoting pattern. 
A fluctuation from the edge of the truncated disc first gives a change in seed photons to the soft Comptonisation. 
Assuming an increase in seed photons, it increases the Compton cooling on the light travel time without any change in electron heating, so the spectrum softens. 
Then, after the accretion time-scale (propagation time-scale), the same fluctuation modulates the soft Comptonisation by increasing the electron density, increasing the heating rate, and causing the spectrum to harden. 
This gives an asynchronous rocking of the soft Comptonisation, where two mutually-correlated but lagged variability sources form its time-dependent behaviour.
By contrast, in the hard Comptonisation region, the fluctuation gives a synchronous change in seed photons and electron heating as both are produced together around its outer edge. 
The synchronous pivoting implemented in our model may be limiting its ability to properly model the data, as it is suppressing a real lag which occurs from the two time-scale propagation mechanism in the soft Comptonisation. 

It is worth noting that our implementation of the spectral pivoting is different from that in the {\tt RELTRANS} model (e.g. \citealt{Mastroserio_2018, Ingram_2019b}).
\cite{Mastroserio_2018, Mastroserio_2019, Mastroserio_2021} consider the nonlinear effects in the time-varying continuum spectrum and have two variable terms in its expression to allow lags to arise from the spectral pivoting itself (\citealt{Kotov_2001}).
However, not specifying the underlying process causing the spectral pivoting may make {\tt RELTRANS} too flexible in producing the observed hard-lag data.
On the other hand, our new model has only one variable term, i.e., the local mass accretion rate, in the expression of the local spectrum.
The local spectrum varies linearly to this term, which does not produce lags except for $\pi$.
As mentioned above, our model relies on the hard lags caused by the combination of the propagating fluctuations process and energy-dependent emission profile (\citealt{Veledina_2016, Veledina_2018}).

We suspect that this lack of a physical spectral pivoting model, including the light crossing time spectral softening as well as the propagation time, is the major reason our new model fails to fit the phase-lag spectra from 50--150~keV.
This more physical model for spectral pivoting is beyond the scope of this paper but will be considered in future work. 


\section{Conclusions}
\label{sec:conclusions}

We have studied X-ray spectral-timing properties of the black hole binary MAXI J1820+070 in the bright low/hard state using {\it Insight-HXMT} observation data.
Particularly, we have focused on the energy-dependent broad-band variability on time-scales from milliseconds to seconds.

We started with testing our previous model from K22, which included the propagating fluctuations process and reverberation and successfully explained soft X-ray timing properties ($< 10\,\si{keV}$), and found that it cannot be applied to higher energy bands.
The key variability feature that our previous model missed was the decrease in fractional power spectrum with energy above $\sim 10\,\si{keV}$, which is difficult to explain with the simple propagating fluctuations picture but is typically observed (\citealt{Nowak_1999, Malzac_2003, Axelsson_2018}).
We have seen that additional effects proposed in the literature, such as the damping (\citealt{Rapisarda_2017a, Mahmoud_2018b}), are not very effective in reproducing both observed power spectra and phase-lag spectrum simultaneously.

We updated our model by implementing spectral pivoting. This is physically expected in a Comptonisation model with fluctuating power (\citealt{Malzac_2003, Gandhi_2008, Veledina_2016, Mastroserio_2018}). This, plus allowing the propagation speed to be different to the time-scale on which fluctuations are generated, allows us to reproduce both power spectra and phase-lag spectra across the 2.6--48~keV band.
The pivoting reduces the amplitude of the response of the spectrum at high energies compared to low energies, suppressing the power spectral normalisation at high energies but keeping its shape. 
We are finally able to do a full spectral-timing fit in the 2--50~keV bandpass, demonstrating that our timing model can be self-consistently combined with spectral models.

The propagation derived from the spectral-timing fit favours SANE than MAD, as the propagation speed is too fast to explain the observed lags for MAD.
Our spectral model for the accretion flow consists of emission from the turbulent disc, plus soft and hard Comptonisation regions and their associated disc reflection. 
These spectral components fit all the emission in our bandpass, and their timing components fit all of the power spectra and phase lags apart from the QPO. 
Thus there is very little room for any additional spectral component to make the QPO, supporting models where the QPOs originate from a global modulation (multiplicative process) of the existing hot flow such as Lense-Thirring precession (\citealt{Ingram_2009}) rather than e.g. an additional component from the jet (\citealt{Ma_2021}).

Nonetheless, our new model still has some limitations, with clear discrepancies with data above $\sim 40\,\si{keV}$. The observed phase lags keep on increasing up to $\sim 150$~keV while those in our model saturate above $\sim 40\,\si{keV}$. These may point to a more complex description of spectral pivoting and/or additional spectral stratification of the inner parts of the hot flow, but clearly, they show that the data are better than the best current models of the hot flow. 
We stress that this is a motivation for better physical modelling of the flow so that we can robustly use the observed spectral and timing data to explore the underlying nature and geometry of the accretion flow in the region where the jet is launched.


\section*{Acknowledgements}

We thank Zi-Xu Yang, Liang Zhang, and Xiang Ma for help with {\it Insight-HXMT} data reduction.
We acknowledge the anonymous referee for their valuable comments.
This work was supported by JSPS KAKENHI Grant Numbers 18H05463 and 20H00153, and WPI MEXT.
TK acknowledges support from JST SPRING, Grant Number JPMJFS2108. 
CD acknowledges support from the STFC consolidated grant ST/T000244/1.
Our work also benefited from discussions during the workshop at the International Space Science Institute (Bern).

We dedicate this work to the memory of Magnus Axelsson, who died suddenly as we were finishing the paper. 

\section*{Data Availability}

The observational data underlying this article are available at \url{http://hxmten.ihep.ac.cn/}. 
The model will be shared upon reasonable request to the corresponding author.



\bibliographystyle{mnras}
\bibliography{manuscript.bib} 



\appendix

\section{Summary of our previous model}
\label{sec:app_model_summary}

We give the detailed formalism of our model in the whole appendix.
We start with the summary of our previous model presented in K22 before formulating model updates made in this work in the subsequent appendices.
We note that while we assume three radially-stratified spectral components as the variable flow, i.e., $N_{\mathrm{s}}=3$ (see Fig.~\ref{fig:propagating_fluctuations_schematic} (a)), the formalism can be straightforwardly applied to an arbitrary number of spectral components.
Needless to say, the formalism is the simplest for the single spectral component, $N_{\mathrm{s}}=1$.

We split the variable flow ranging from $r_{\mathrm{in}}$ to $r_{\mathrm{out}}$ into $N_{\mathrm{r}}$ rings logarithmically, such that the central radius of the $n$th ring, $r_{n}$, and the distance between neighboring central radii, $\Delta r_{n}=r_{n} - r_{n+1}$, follow $\Delta r_{n}/r_{n}=\mathrm{constant}$ ($n=1, 2, \cdots, N_{\mathrm{r}}$ from outer to inner rings).
We use the radius $r_{n}$ in units of gravitational radii $R_{\mathrm{g}}=GM_{\mathrm{BH}}/c^{2}$, where $G,\,M_{\mathrm{BH}},\,c$ are the gravitational constant, black hole mass, and speed of light in vacuum, respectively.
The number of rings $N_{\mathrm{r}}$ determines the resolution of the model calculation (\citealt{Ingram_2013}).
The larger number of $N_{\mathrm{r}}$ allows more accurate calculations of power spectra and cross spectra for high frequencies at the expense of computation efficiency.

The mass accretion rate varies stochastically everywhere in the variable flow.
We define this intrinsically fluctuating mass accretion rate for each ring as the product of its mean $m_0$ and time-variable term $a(r_n, t)$ having the mean of unity, $\mu=\langle a(r_n, t) \rangle _t =1$, where $t$ denotes time.
Employing a dimensionless mass accretion rate prescription, we set $\dot{m}_{0}=1$.
We assume that the power spectrum of $a (r_n, t)$, $|A(r_n, f)|^2$, is provided by a zero-centred Lorentzian function with the cut-off frequency equal to the generator frequency $f_{\mathrm{gen}}(r_n)$,
\begin{equation}
    |A(r_n, f)|^2=\frac{2 \sigma ^2}{\pi \mu ^2} \frac{f_{\mathrm{gen}}(r_n)}{f^2 + (f_{\mathrm{gen}}(r_n))^2},
    \label{eq:mdot_power_intr}
\end{equation}
where $f$ is the Fourier frequency and $\sigma ^2$ the variance of $a(r_n, t)$.
In the expression (\ref{eq:mdot_power_intr}), we employ the normalised power spectra such that their integral over positive frequency corresponds to $(\sigma/\mu)^2$.
We use the parameter $F_\mathrm{var}$ to set the variance through $\sigma / \mu = F_{\mathrm{var}}/\sqrt{N_{\mathrm{dec}}}$, where $N_{\mathrm{dec}}$ is the number of rings per radial decade and thus, $F_\mathrm{var}$ is the fractional variability per radial decade.
The generator frequency is defined as equation (\ref{eq:viscous_frequency}) (see also Fig.~\ref{fig:propagating_fluctuations_schematic} (c)).
Sample power spectra are shown in Fig.~\ref{fig:propagating_fluctuations_schematic} (e) with dashed lines.

While mass accretion rate fluctuations are generated, they propagate towards the central object at the same time.
The propagation is expected to happen in a multiplicative manner (\citealt{Uttley_2005}), which leads to assuming 
\begin{equation}
    \dot{m}(r_n, t)=\dot{m}_0 \prod _{k=1} ^{n} a(r_{k}, t-\Delta t_{k, n}),
\end{equation}
where $\Delta t_{k, n}$ is the propagation time from the outer $k$th ring to inner $n$th ($n \geq k$) ring.
We note that $\dot{m}_0 =1$ is left in the expression to preserve generality.
As the propagation speed $v_{\mathrm{p}}(r)$ is set by radial velocity, i.e., $v_{\mathrm{p}}(r)=r f_{\mathrm{\mathrm{visc}}}(r)$, the propagation time is expressed as 
\begin{equation}
	\Delta t_{k, n}=\frac{\Delta r}{r} \sum _{l=k+1} ^{n} \frac{1}{f_{\mathrm{visc}}(r)},
	\label{eq:prop_time}
\end{equation}
where $\Delta r/r=\Delta r_{n}/r_{n}=\mathrm{constant}$.
In the statistical equilibrium under the generation and propagation of mass accretion rate fluctuations, the power spectra, $|\dot{M}(r_n, f)|^2$, and cross spectra, $(\dot{M}(r_m, f))^{*} \dot{M}(r_n, f) \,(m \neq n)$, can be calculated analytically (see Appendix A in K22).
The asterisk $*$ denotes the complex conjugate.
Sample the power spectra are shown in Fig.~\ref{fig:propagating_fluctuations_schematic} (e) with solid lines.

The flux at energy $E$ and time $t$, $x(E, t)$, is the sum of contributions from each ring in the variable flow.
Ignoring the effect of reverberation for simplicity (we take it into account in Section \ref{sec:app_reflection}), we express the flux as 
\begin{equation}
	x(E, t)=\sum _{n=1} ^{N_{\mathrm{r}}} \lambda (r_{n}) S(E, r_{n}, t).
	\label{eq:flux_total}
\end{equation}
The flux represents the number of photons per unit time because the variability is studied for count rather than energy.
Both treatments give the same model calculations for a single energy.
However, using count rates is necessary to compare the model to observations because observation data cannot be processed for an infinitesimal energy band.
The local flux emitted from $n$th ring, $S(E, r_{n}, t)$, is defined as 
\begin{equation}
	S(E, r_{n}, t)=S_{0}(E, r_{n}) \frac{\dot{m}(r_n, t)}{\dot{m}_{0}},
	\label{eq:flux_local_pre}
\end{equation}
where $S_{0}(E, r_{n})$ is the time-averaged spectrum at the $n$th ring.
We note that $\dot{m}_0 =1$ is explicitly written in equation (\ref{eq:flux_local_pre}).
Since we assume three spectral components ($N_{\mathrm{s}}=3$), this time-averaged spectrum is categorised into three spectra:
\begin{equation}
    S_{0} (E, r_n)=
    \begin{cases}
        S_{\mathrm{h}}(E) & (r_{\mathrm{in}} \leq r_{n} < r_{\mathrm{sh}})  ,\\
        S_{\mathrm{s}}(E) & (r_{\mathrm{sh}} \leq r_{n} < r_{\mathrm{ds}})  ,\\
        S_{\mathrm{d}}(E) & (r_{\mathrm{ds}} \leq r_{n} < r_{\mathrm{out}}) ,
    \end{cases}
\end{equation}
depending on to which spectral regions the $n$th ring belongs.
$r_{\mathrm{in}}$ is the inner edge of the hard Comptonisation component, $r_{\mathrm{sh}}$ is the transition radius between the hard Comptonisation and soft Comptonisation components, $r_{\mathrm{ds}}$ is the transition radius between the soft Comptonisation and variable disc components, and $r_{\mathrm{out}}$ is the outer edge of the variable disc component.
$S_{\mathrm{h}}(E),\,S_{\mathrm{s}}(E),\,S_{\mathrm{d}}(E)$ are the time-averaged spectra for the hard Comptonisation, soft Comptonisation, and disc components, as shown in Fig.~\ref{fig:propagating_fluctuations_schematic} (d).

To let equations be concise, we define the collection of radii affiliated with the variable disc, soft Comptonisation, and hard Comptonisation regions as $\vb*{r}_{\mathrm{d}}$, $\vb*{r}_{\mathrm{s}}$, and $\vb*{r}_{\mathrm{h}}$, respectively.
Then we express $S_{0}(E)$ as 
\begin{equation}
	S_{0} (E, r_n)=S_{\mathrm{Y}}(E)\,\,(r_n \in \vb*{r}_{\mathrm{Y}}),
\end{equation}
for $\mathrm{Y}=\mathrm{h, s, d}$.
In the model formalism, we use $S_{0}(E, r_{n})$ as a fraction, giving the constraint of
\begin{equation}
    \sum _{\mathrm{Y}=\mathrm{h, s, d}} S_{\mathrm{Y}}(E)=1.
    \label{eq:spec_sum}
\end{equation}
We note that we ignore any other spectral components here, such as reflected emission.

Along with the time-averaged spectrum $S_{0}(E, r_{n})$, the contribution from each ring is also regulated by $\lambda (r_{n})$ in terms of energy dissipation:
\begin{equation}
	\lambda (r_{n}) = \epsilon (r_{n}) 2\pi r_{n} \Delta r_{n} \left/ \sum _{m\,(r_m \in \vb*{r}_{\mathrm{Y}})} \epsilon (r_{m}) 2\pi r_{m} \Delta r_{m} \right. ,
	\label{eq:lambda}
\end{equation}
where $\epsilon (r)$ is the emissivity.
The emissivity is defined as the product of a power-law function and inner boundary condition: $\epsilon (r) \propto r^{-\gamma}b(r)$, where $b(r)=1-\sqrt{r_{\mathrm{in}}/r}$ or $b(r)=1$ for the `stress-free' or `stressed' boundary condition (\citealt{Ingram_2012}). 
Due to the normalisation of $\lambda (r_{n})$ and mass accretion rate fluctuations, the time-averaged flux from an entire spectral region corresponds to the time-averaged spectrum:
\begin{equation}
	\sum _{n\,(r_n \in \vb*{r}_{\mathrm{Y}})} \lambda (r_{n}) S(E, r_{n}, t)=S_{\mathrm{Y}}(E),
\end{equation}
giving rise to the time-averaged total flux being unity, $\langle x(E, t) \rangle _{t}=\sum _{\mathrm{Y}=\mathrm{h, s ,d}}S_{\mathrm{Y}}(E)=1$.

From equations (\ref{eq:flux_total}) and (\ref{eq:flux_local_pre}), the Fourier transform of the flux $x(E, t)$ is proportional to the Fourier transform of the local mass accretion rate $\dot{m}(r_n, t)$:
\begin{equation}
    \begin{split}
	X(E, f) &= \sum _{n=1} ^{N_{\mathrm{r}}} \lambda (r_{n}) \frac{S_{0}(E, r_{n})}{\dot{m}_0} \dot{M}(r_n, f)\\
	&=\sum _{n=1} ^{N_{\mathrm{r}}} w(r_n, E) \dot{M}(r_n, f),
	\end{split}
\label{eq:x_ft_pre}
\end{equation}
where we call the coefficient $w(r_n, E)=\lambda (r_n)(S_{0} (E, r_n)/\dot{m}_0)$ the weight.
The weight $w(r_n, E)$ is proportional to the product of the emissivity and energy spectrum, finally determining how much $n$th ring contributes to the variability of the flux.
Since we know the analytic forms of power spectra and cross spectra for the local mass accretion rate, $(\dot{M}(r_m, t))^{*} \dot{M}(r_n, t)\,(m, n=1, 2, \cdots, N_{\mathrm{r}})$, we can calculate the power spectrum $|X(E, f)| ^2$ and cross spectrum for two different energies $E_{1}$ and $E_{2}$, $(X(E_{1}, f))^{*} X(E_{2}, f)$, analytically (see Appendix A in K22), which can be directly compared to observation data.
The sample model calculations of power spectra, time-lag spectrum, and phase-lag spectrum are shown in  Fig.~\ref{fig:propagating_fluctuations_schematic} (f), (g), and (h), respectively.


\section{Including damping effects}
\label{sec:app_damping}

We summarize the expressions of power spectra and cross spectra for the local mass accretion rate and flux, where the damping effects are taken into account.
The derivations are described in \cite{Rapisarda_2017a} Appendix A and in \cite{Mahmoud_2018b} Appendix A.
We note that the damping effects are excluded in any results presented since they turned out to be ineffective for this work. 

The local mass accretion rate is expressed as
\begin{equation}
    \dot{m}(r_n, t)=\prod _{k=1} ^{n} g(r_k, r_n, t) \otimes a (r_k, t),
    \label{eq:mdot_general}
\end{equation}
where the symbol $\otimes$ denotes the convolution in time $t$.
The green function $g(r_k, r_n, t)$ describes how the local mass accretion rate at the inner $n$th ring responds to that at the outer $k$th ring ($n \geq k$).
The Fourier transform of the green function, $G(r_k, r_n, t)$, is required in model calculations.
While we employ $G(r_k, r_n, t)=\mathrm{e}^{-i2\pi f\Delta t_{k, n}}$ in K22 and Appendix~\ref{sec:app_model_summary}, which is equivalent to $g(r_k, r_n, t)=\delta (t-\Delta t_{k, n})$, we use 
\begin{equation}
    G(r_k, r_n, f)=\mathrm{e}^{-D f \Delta t_{k, n}}\mathrm{e}^{-i2 \pi f \Delta t_{k, n}}
\end{equation}
in order to include the damping effects.

From equation (\ref{eq:mdot_general}), the following expressions for power spectra and cross spectra for the local mass accretion rate are obtained:
\begin{equation}
    |\dot{M}(r_n, f)|^{2}=\frac{1}{N^2}|A(r_n, f)|^2 \otimes | G(r_{n-1}, r_n, f)\dot{M}(r_{n-1}, f)|^2,
    \label{eq:mdot_power_general}
\end{equation}
\begin{equation}
    (\dot{M}(r_k, f))^{*} \dot{M}(r_n, f) =\Lambda _{k,n} G(r_{k}, r_n, f)|\dot{M}(r_{k}, f)|^2 \,\,\,(n \geq k),
    \label{eq:mdot_cross_general}
\end{equation}
where $\otimes$ denotes the convolution in frequency $f$.
The consecutive product of the time-averaged local mass accretion rate $\mu_{l}$ ($l=1, \cdots, N_{\mathrm{r}}$) from $k$th ring to $n$th ring is defined as $\Lambda _{k, n}$, i.e., $\Lambda _{k, n}=\prod _{l=k+1} ^{n} \mu_{l}$, which is unity.
We note that we use continuous Fourier frequency $f$ rather than discrete Fourier frequency $f_{m}$ for simplicity.
The coefficient $1/N^2$ in equation (\ref{eq:mdot_power_general}), where $N$ is the number of data points used, comes from the convolution theorem 
\begin{equation}
    X (f_{m}) = \frac{1}{N} A(f_m) \otimes B(f_m),
\end{equation}
where $x(t_{l})=a(t_{l})b(t_{l})$.
In the convolution theorem, the discrete Fourier transform is defined as 
\begin{equation}
    X (f_m) = \sum _{l=0} ^{N-1} x (t_{l})\mathrm{e}^{-i 2 \pi f_m t_l},
\end{equation}
where $f_m =m/(N\Delta t)$ ($m=-N/2+1, \cdots, N/2$ for even $N$ or $m=(N-1)/2, \cdots, (N-1)/2$ for odd $N$) and $t_l =l \Delta t$ with $\Delta t$ being the sampling interval.
The discrete inverse Fourier transform is thus 
\begin{equation}
    x (t_l) = \frac{1}{N} \sum _{m} X (f_{m})\mathrm{e}^{+i 2 \pi f_m t_l}.
\end{equation}

Combining equations (\ref{eq:mdot_power_general}), (\ref{eq:mdot_cross_general}) with equation (\ref{eq:weight}) yields the expressions of the power spectra for the flux
\begin{equation}
    \begin{split}
    |X(E, f)|^2 =& \sum _{n=1} ^{N_{\mathrm{r}}}  \biggl[ (w(r_n, E))^2 |\dot{M}(r_n, f)|^2 \\ 
    & + 2\sum _{k=1} ^{n-1} \Bigl\{ w(r_k, E)w(r_n, E) \Lambda _{k,n}\\
    & \times |G(r_{k}, r_n, f)|\cos (\Phi (r_k, r_n, f))|\dot{M}(r_{k}, f)|^2 \Bigr\} \biggr],
    \end{split}
\end{equation}
where $G(r_k, r_n, f)=|G(r_k, r_n, f)|\mathrm{e}^{i\Phi (r_k, r_n, f)}$.
The cross spectra for the flux can be calculated in the same manner:
\begin{equation}
    \begin{split}
    (X(E_1, f))^{*}X(E_2, f)=& \sum _{n=1} ^{N_{\mathrm{r}}}  \biggl[ w(r_n, E_1)w(r_n, E_2) |\dot{M}(r_n, f)|^2 \\ 
    & + \sum _{k=1} ^{n-1} \Bigl\{ \bigl( w(r_k, E_1)w(r_n, E_2)\mathrm{e}^{-i\Phi(r_k, r_n, f)}\\ 
    &+ w(r_n, E_1)w(r_k, E_2)\mathrm{e}^{+i\Phi(r_k, r_n, f)}\bigr) \\
    & \times \Lambda _{k,n} |G(r_{k}, r_n, f)| |\dot{M}(r_{k}, f)|^2 \Bigr\} \biggr].
    \end{split}
\end{equation}


\section{Including spectral pivoting}
\label{sec:app_pivot}

To include the change in the spectral shape on short time-scales, we redefine the variable local energy spectrum $S(E, r_n, t)$ as 
\begin{equation}
	S(E, r_{n}, t) = \left( 1+\eta (E, r_{n}) \frac{\dot{m}(r_{n}, t)-\dot{m_{0}}}{\dot{m_{0}}} \right) S_{0}(E, r_{n}).
	\label{eq:flux_local_new}
\end{equation}
Again, although we assume $\dot{m}_0=1$, we leave the symbol explicitly in the formulation to preserve generality.
The new energy-dependent parameter $\eta (E, r_{n})$ regulates how the spectrum at the energy $E$ responds to mass accretion rate fluctuations.
While the amplitude of $\eta (E, r_n)$ regulates the sensitivity, its sign determines the correlation patterns (see Fig.~\ref{fig:schematic_spectral_pivot}).
Since we have three spectral regions for the variable flow ($N_{\mathrm{s}}=3$), we accordingly classify this sensitivity parameter into three:
\begin{equation}
    \eta (E, r_n)=\eta _{\mathrm{Y}}(E)\,\,(r_n \in \vb*{r}_{\mathrm{Y}}),
\end{equation}
for $\mathrm{Y}=\mathrm{h, s, d}$.
The spectrum $S(E, r_n, t)$ is positively (negatively) correlated to mass accretion rate fluctuations if $\eta(E, r_{n})>0$ ($<0$), as shown in Fig.~\ref{fig:schematic_spectral_pivot}.
Both correlations can be physically realized (\citealt{Veledina_2018}).
The new definition (\ref{eq:flux_local_new}) is reduced to the previous definition (\ref{eq:flux_local_pre}) if $\eta (E, r_{n})=1$.
Thus, we see that equation (\ref{eq:flux_local_new}), in which the spectral shapes vary in time, is a natural extension of (\ref{eq:flux_local_pre}), in which the spectral shapes are fixed.

Employing equation (\ref{eq:flux_local_new}), the Fourier transform of the flux is  
\begin{equation}
	\begin{split}
		X(E, f) =& \sum _{n=1} ^{N_{\mathrm{r}}} \lambda (r_{n}) \frac{ \eta(E, r_{n}) S_{0}(E, r_{n})}{\dot{m}_0} \dot{M}(r_n, f) \\
		=&\sum _{n=1} ^{N_{\mathrm{r}}} w(r_{n}, E) \dot{M}(r_{n}, f).
	\end{split}
	\label{eq:x_ft}
\end{equation}
where the weight $w(r_{n}, E)$ is redefined as
\begin{equation}
	w(r_{n}, E) = \lambda (r_{n}) \frac{\eta (E, r_{n}) S_{0}(E, r_{n})}{\dot{m}_{0}}.
	\label{eq:weight}
\end{equation}
The Fourier transform $X(E, f)$ is expressed in the same manner as equation (\ref{eq:x_ft_pre}) and thus linear to $\dot{M}(r_{n}, f)$.
Thus, we can calculate power spectra and cross spectra for the flux analytically. 

The only difference that appeared in the expression of $X(E, f)$ (compare equations (\ref{eq:x_ft_pre}) and (\ref{eq:x_ft})) is the presence of the new parameter $\eta (E, r_{n})$.
Allowing the spectral variation is equivalent to regarding $S_{0}(E)$ in equation (\ref{eq:x_ft_pre}) as $\eta (E, r_{n})S_{0}(E, r_{n})$.
Due to this replacement, the spectral pivoting releases the model from the constraint (\ref{eq:spec_sum}) in the sense that the model calculations always require the product $\eta (E, r_n)S(E, r_n)$, not $S_{0}(E, r_n)$. 
The spectral pivoting gives much flexibility to the model in this way, as seen in the successful modelling in Fig.~\ref{fig:psd_phase_fit_comp} (right-most) compared to the failed ones in the rest of the columns.


\section{Modifying reverberation}
\label{sec:app_reflection}

We modify the implementation of reflection due to the implementation of the spectral variation described above.
While a part of the emission from the variable flow directly hits a detector, it also irradiates the outer disc, resulting in the reflected/reprocessed emission.
Considering this reprocessed emission also hits the detector, the observed flux is expressed as 
\begin{equation}
	x(E, t)=x^{(\mathrm{d})}(E, t) + x^{(\mathrm{r})}(E, t),
	\label{eq:flux_total_ref}
\end{equation}
where $x^{(\mathrm{d})}(E, t)$ and $x^{(\mathrm{r})}(E, t)$ are the direct and reflected components, respectively.
Hereafter, we use the subscripts `$(\mathrm{d})$' and `$(\mathrm{r})$' to specify the direct and reflected components.
Since we have only considered the direct component above, symbols that appeared in previous appendices will be used with the upper subscript of `$(\mathrm{d})$'.
The expression of the direct component $x^{(\mathrm{d})}(E, t)$ corresponds to equation (\ref{eq:flux_total}).

The variable disc component is unlikely to contribute to the reflected emission.
In this regard, the variable disc component is distinct from the soft and hard Comptonisation components.
However, in terms of formalism, it is simple to assume that every spectral component in the variable flow, i.e. every direct component, has its associated reflected component.
Therefore, we account for the reflected component for the variable disc in the formalism and remove it in parameter space.

Each ring of the variable flow illuminates the outer disc, yielding the following expression of the reflected component:
\begin{equation}
x ^{(\mathrm{r})}(E, t)=\sum _{n=1} ^{N_{\mathrm{r}}} \int dE' \lambda (r_n)S^{(\mathrm{d})}(E', r_n, t) \otimes h(E, E', r_n, t),
\label{eq:flux_total_reflection}
\end{equation}
where the symbol $\otimes$ denotes the convolution in time $t$.
The impulse response $h(E, E', r_n, t)$ describes the time-evolution of reflected flux at the photon energy $E$ for the instant incident flux at $E'$ from the $n$th ring.
The impulse response also encompasses the probability that a photon at energy $E$ is generated from one at $E'$.
Here, we assume that the form of the impulse response is common within each spectral component: 
\begin{equation}
    h (E, E', r_n, t)=h _{\mathrm{Y}}(E, E', t)\,\,(r_n \in \vb*{r}_{\mathrm{Y}}),
    \label{eq:imp_resp_spec_comp}
\end{equation}
for $\mathrm{Y}=\mathrm{h, s, d}$.
The radius of the ring $r_{n}$ is only needed to specify to which spectral component the ring belongs.
The impulse response for each spectral component is independent of radius, i.e., the location of illuminating source.
This approximation should be validated since the size of a reflector ($\sim 10^{3} R_{\mathrm{g}}$) is expected to be much larger than the size of spectral regions in the variable flow ($\sim 10 R_{\mathrm{g}}$).

We also assume that the shape of the impulse response is independent of photon energies $E$ and $E'$ and separate its amplitude and shape in the expression:
\begin{equation}
h(E, E', r_n, t)=C(E, E', r_n)\widetilde{h}(r_n, t),
\label{eq:imp_resp_separation}
\end{equation}
where we normalise the shape $\widetilde{h}(t)$ to follow 
\begin{equation}
\int _{-\infty} ^{+\infty} dt \,\widetilde{h}(r_n, t)=1.
\label{eq:imp_resp_norm_condition}
\end{equation}
Again, each term of the impulse response is common within spectral regions:
\begin{equation}
    C(E, E', r_n)=C _{\mathrm{Y}}(E, E'),\, 
    \widetilde{h}(r_n, t)=\widetilde{h}_{\mathrm{Y}}(t) \,\,(r_n \in \vb*{r}_{\mathrm{Y}}),
\end{equation}
for $\mathrm{Y}=\mathrm{h, s, d}$.
Hereafter, we replace the symbol of the normalised impulse response $\widetilde{h}(r_n, t)$ with $h(r_n, t)$, since the symbol of the impulse response including normalisation, $h(E, E', r_n, t)$, is not required in the formalism anymore.
Substituting equations (\ref{eq:flux_local_new}) and (\ref{eq:imp_resp_separation}) into equation (\ref{eq:flux_total_reflection}) yields
\begin{equation}
\begin{split}
x ^{(\mathrm{r})}(E, t)=&\sum _{n=1} ^{N_{\mathrm{r}}} \int dE' \lambda (r_n) \left( 1+\eta ^{(\mathrm{d})} (E', r_n) \frac{\dot{m}(r_{n}, t)-\dot{m_{0}}}{\dot{m_{0}}} \right) \\
&\times S^{(\mathrm{d})} _{0}(E', r_n) \otimes C(E, E', r_n)h(r_n, t).    
\end{split}
\label{eq:flux_total_reflection_a}
\end{equation}
We note that $(\mathrm{d})$ is explicitly used here to distinguish the direct component from the reflected one, although not in the previous subsections, where the reflection is not included.

To connect equation (\ref{eq:flux_total_reflection_a}) with the time-averaged reflected spectrum
\begin{equation}
S_0 ^{(\mathrm{r})}(E, r_n)= S ^{(\mathrm{r})} _{\mathrm{Y}}(E)\,\,(\mathrm{Y}=\mathrm{h,\,s,\,d}),    
\end{equation}
we average the reflected flux $x ^{(\mathrm{r})}(E, t)$ in time:
\begin{equation}
\langle x ^{(\mathrm{r})}(E, t) \rangle _{t}=\sum _{\mathrm{Y}=\mathrm{h, s, d}} \int dE' C_{\mathrm{Y}}(E, E') S^{(\mathrm{d})} _{\mathrm{Y}}(E'),
\label{eq:flux_total_reflection_average}
\end{equation}
which needs to correspond to $\sum _{\mathrm{Y}=\mathrm{h, s, d}} S_{\mathrm{Y}} ^{(\mathrm{r})}(E)$.
Each term on the right-hand side corresponds to each time-averaged reflected flux.
In practical applications, we fix $S^{(\mathrm{r})}_{\mathrm{d}}(E)=0$ to exclude the reflected emission arising from irradiation by the variable disc component.
Equation (\ref{eq:flux_total_reflection_a}) cannot be simplified with this constraint because $\eta ^{(\mathrm{d})}(E', r_n)$ depends on the incident photon energy $E'$.
The existence of $\eta ^{(\mathrm{d})}(E', r_n)$ in the integration in equation (\ref{eq:flux_total_reflection_a}) means that the spectral variation at every energy affects the reflected emission as long as $C(E, E', r_n) \neq 0$.
Although this situation is physically natural, its proper treatment would require many complications.
Moreover, $\eta ^{(\mathrm{d})} (E', r_n)$ is an \textit{ad hoc} parameter introduced to include the spectral variation and not connected directly to physical parameters.
Given these circumstances, we replace $\eta ^{(\mathrm{d})} (E', r_n)$ with $\eta ^{(\mathrm{r})} (E, r_n)$ related to the reflection component and let $\eta ^{(\mathrm{r})} (E, r_n)$ be only dependent of the output photon energy $E$, resulting in 
\begin{equation}
    \begin{split}
    x ^{(\mathrm{r})}(E, t)&=\sum _{n=1} ^{N_{\mathrm{r}}} \lambda (r_n) \left( 1+\eta ^{(\mathrm{r})} (E, r_n) \frac{\dot{m}(r_{n}, t)-\dot{m_{0}}}{\dot{m_{0}}} \right)\\
    & \times S^{(\mathrm{r})} _{0}(E, r_n) \otimes h(r_n, t).    
    \end{split}
\label{eq:flux_total_reflection_b}
\end{equation}
Defining 
\begin{equation}
S^{(\mathrm{r})} (E, r_{n}, t) =\left( 1+\eta ^{(\mathrm{r})} (E, r_n) \frac{\dot{m}(r_{n}, t)-\dot{m_{0}}}{\dot{m_{0}}} \right) S^{(\mathrm{r})}(E, r_n)
\label{eq:flux_local_reflection_a}
\end{equation}
yields an expression for the reflected component comparable to that for the direct component (\ref{eq:flux_total}):
\begin{equation}
x ^{(\mathrm{r})}(E, t)=\sum _{n=1} ^{N_{\mathrm{r}}} \lambda (r_n)  S^{(\mathrm{r})}(E, r_n, t) \otimes h(r_n, t).
\label{eq:flux_total_reflection_c}
\end{equation}
The difference lies in the presence of convolution.
As each direct spectral component does, each reflected spectral component shares the sensitivity parameter within its associated radii:
\begin{equation}
	\eta ^{(\mathrm{r})}(E, r_n)=\eta _{\mathrm{Y}}(E)\,\,(r_n \in \vb*{r}_{\mathrm{Y}})
\end{equation}
for $\mathrm{Y}=\mathrm{h, s, d}$.
In the spectral-timing fit in Section~\ref{sec:spec_timing}, we assume $\eta ^{(\mathrm{d})}_{\mathrm{Y}}(E)=\eta ^{(\mathrm{r})} _{\mathrm{Y}}(E)$, which means that the direct and its associated reflected components react to a deviation of the mass accretion rate from its average in the same way.

Finally, substituting equations (\ref{eq:flux_total}) and (\ref{eq:flux_total_reflection_c}) into equation (\ref{eq:flux_total_ref}), we get the expression of the flux:
\begin{equation}
	x(E, t)=\sum _{n=1} ^{N_{\mathrm{r}}} \lambda (r_n) \left( S^{(\mathrm{d})}(E, r_n, t) + S^{(\mathrm{r})}(E, r_n, t) \otimes h(r_n, t) \right) .
	\label{eq:flux_total_ref_a}
\end{equation}
Using equations (\ref{eq:flux_local_new}) and (\ref{eq:flux_local_reflection_a}), its Fourier transform is 
\begin{equation}
	X(E, f)=\sum _{n=1} ^{N_{\mathrm{r}}} \left( w^{(\mathrm{d})}(r_n, E) +  w^{(\mathrm{r})}(r_n, E)H(r_n, f) \right) \dot{M}(r_{n}, f),
	\label{eq:fourier_flux_total_ref}
\end{equation}
where
\begin{equation}
    w^{(\mathrm{r})}(r_n, E)= \lambda (r_n) \frac{\eta ^{(\mathrm{r})}(E, r_n)S^{(\mathrm{r})} _{0} (E, r_n)}{\dot{m}_0}.
\end{equation}
is the weight for the reflection component, and $H(r_n, f)$ is the transfer function, the Fourier transform of the impulse response $h(r_n, t)$.
Again, the linearity of $X(E, f)$ to $\dot{M}(r_{n}, f)$ allows analytic calculations for the power spectra and cross spectra of the flux.


\section{MCMC fit}
\label{sec:app_mcmc}

We demand 28 free parameters in the spectral-timing fit performed in Section~\ref{sec:spec_timing}.
In addition to these many free parameters, our model computations are quite expensive, even though not prohibitive.
These situations make it difficult to properly estimate the errors on the derived parameter values.

Here, we perform an MCMC fit in order to evaluate the uncertainties of the model parameter values.
We use the {\tt chain} command in {\tt XSPEC}, in which we employ the Goodman-Weare algorithm with 100 walkers and 100000 steps in total.
Whereas in Section~\ref{sec:spec_timing}, we free the normalisations of the {\tt lorentz} model used to capture QPO features in the power spectra, we fix all of them in the MCMC fit.
Fixing them decreases 12 free parameters because there are two {\tt lorentz} models for each of the six energy bands, enabling the MCMC analysis with 16 free parameters.
We burn 1000 steps because the $\chi ^2$ values converge after $\sim 1000$ steps.

\begin{figure*}
	\includegraphics[width=\linewidth]{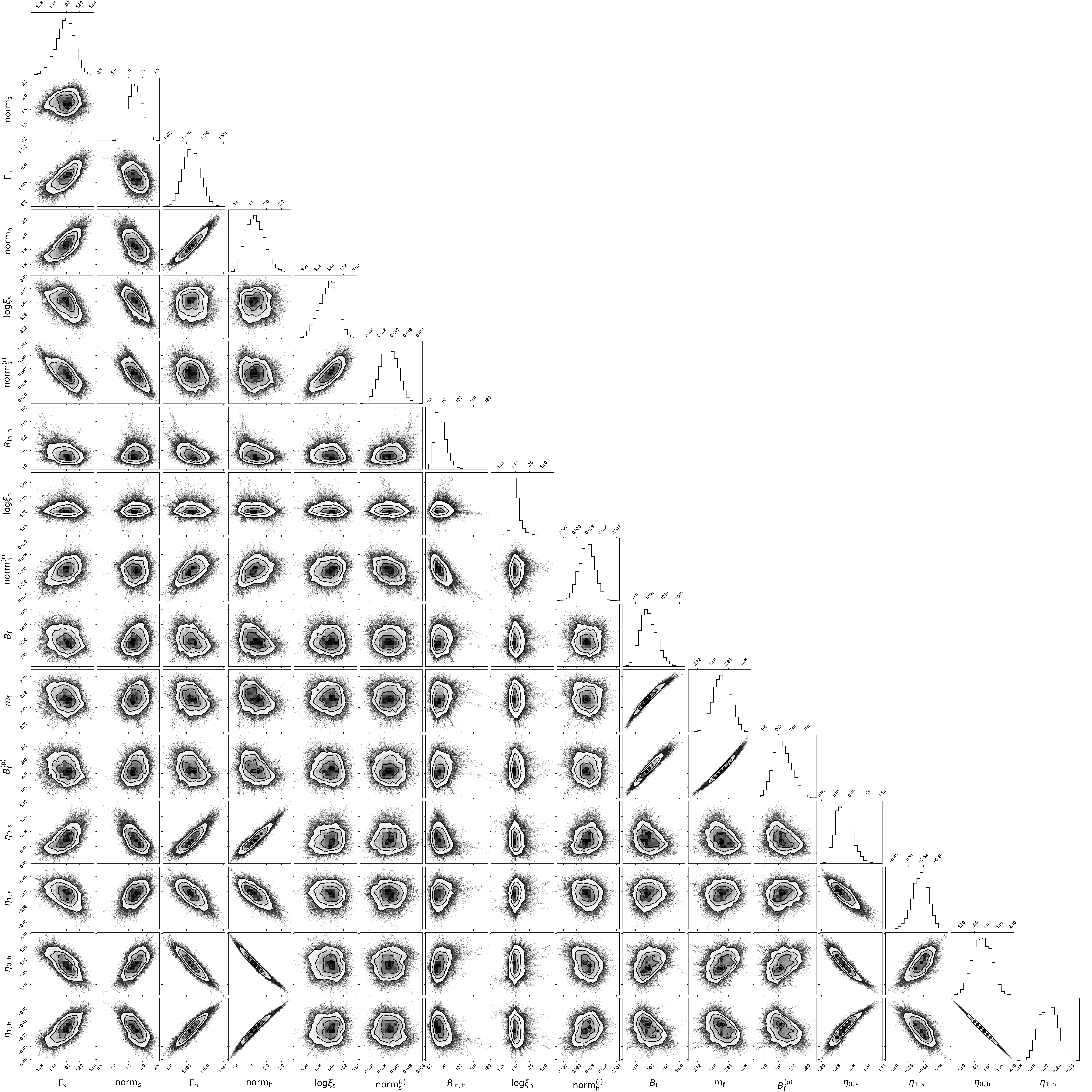}
	\caption{Corner plots for 16 free parameters of the spectral-timing fit in Section~\ref{sec:spec_timing}.
	Every free parameter has its own row and column.
	The correlation between two free parameters is shown at the intersection of the corresponding row and column.
	The distribution of each parameter is shown at the right-most (top) of the corresponding row (column).
	We use the \texttt{corner} module (\citealt{Foreman-Mackey_2016}) to produce the plots.
	The figure is viewed clearly in the digital version.
	}
	\label{fig:mcmc}
\end{figure*}

The corner plots for all 16 free parameters are shown in Fig.~\ref{fig:mcmc}.
The distribution is generally centrally-peaked for every free parameter, signifying good convergence.
The parameters regulating the propagation time-scale, $B^{(\mathrm{p})}_{\mathrm{f}}$ and $m^{(\mathrm{p})}_{\mathrm{f}}$, take values among $160\textrm{--}240$ and $2.78\textrm{--}2.88$, respectively.
These spreads give rise to the spreads of the propagation time-scale as narrow as the line width in Fig.~\ref{fig:fvisc_updates}.
Thus, the result of our analysis that the propagation time-scale derived fairly disagrees with any theoretical predictions in Fig.~\ref{fig:fvisc_updates} is robust.

\bsp	
\label{lastpage}
\end{document}